\documentclass[10pt]{article}
\usepackage{latexsym}
\usepackage{eufrak}
\usepackage{euscript}
\usepackage{graphicx}

\title{ The GRA Beam-Splitter Experiments and Particle-Wave Duality of Light}

\author{ P.N. Kaloyerou\\Deptartment of Physics, School of Natural Sciences,\\ University of Zambia, PO Box 32379, Lusaka 10101, Zambia \footnote{email address: pan.kaloyerou@wolfson.ox.ac.uk}$\;$\footnote{Alternative address: The University of Oxford, Wolfson College, Linton Road,Oxford OX2 6UD, UK.}}

\newcommand{\Ab}{\mbox{\boldmath{$A$}}}
\newcommand{\Abs}{\mbox{{\scriptsize\boldmath{$A$}}}}
\newcommand{\Pb}{\mbox{{ \boldmath{$\mit \Pi$}}}}
\newcommand{\xb}{\mbox{{\scriptsize\boldmath{$x$}}}}
\newcommand{\kb}{\mbox{{\scriptsize\boldmath{$k$}}}}
\newcommand{\kbp}{\mbox{{\scriptsize\boldmath{$k'$}}}}
\newcommand{\rvect}{\mbox{{ \boldmath{$\mit r$}}}}
\newcommand{\Th}{{\mit \Theta}}
\newcommand{\Fi}{{\mit \Phi}}
\newcommand{\fA}{{\mit \Phi}[\Ab,t]}

\newcommand{\ep}{e^{i\kb.\xb}}
\newcommand{\epm}{e^{-i\kb.\xb}}
\newcommand{\akk}{\hat{a}_{k\mu}}
\newcommand{\akkd}{\hat{a}_{k\mu}^{\dagger}}
\newcommand{\qk}{q_{k\mu}}
\newcommand{\qks}{q_{k\mu}^{*}}
\newcommand{\pk}{\pi_{k\mu}}
\newcommand{\pks}{\pi_{k\mu}^{*}}

\newcommand{\ek}{\mbox{\boldmath $\hat\varepsilon$}_{k\mu}}
\newcommand{\ekp}{\mbox{\boldmath $\hat\varepsilon$}_{k'\mu'}}
\newcommand{\V}{V^{\frac{1}{2}}}

\newcommand{\E}{\mbox{\boldmath $E$}}
\newcommand{\Eh}{\mbox{\boldmath $\hat E$}}
\newcommand{\B}{\mbox{\boldmath $B$}}
\newcommand{\I}{\mbox{\boldmath $I$}}
\newcommand{\Ih}{\mbox{\boldmath $\hat I$}}
\newcommand{\epp}{e^{i(\kb-\kbp).\xb}}
\newcommand{\eppm}{e^{-i(\kb-\kbp).\xb}}
\newcommand{\ak}{\alpha_{k_\alpha\mu_\alpha}}
\newcommand{\aks}{\alpha_{k_\alpha\mu_\alpha}^{*}}
\newcommand{\bk}{\beta_{k_\beta\mu_\beta}}
\newcommand{\bks}{\beta_{k_\beta\mu_\beta}^{*}}

\newcommand{\eka}{\mbox{\boldmath $\hat\varepsilon$}_{k_\alpha\mu_\alpha}}
\newcommand{\ekb}{\mbox{\boldmath $\hat\varepsilon$}_{k_\beta\mu_\beta}}
\newcommand{\ekc}{\mbox{\boldmath $\hat\varepsilon$}_{k_c\mu_c}}
\newcommand{\ekd}{\mbox{\boldmath $\hat\varepsilon$}_{k_d\mu_d}}
\newcommand{\eko}{\mbox{\boldmath $\hat\varepsilon$}_{k_0\mu_0}}
\newcommand{\kk}{\mbox{\boldmath $k$}}
\newcommand{\ka}{\mbox{\boldmath $k$}_\alpha}
\newcommand{\kkb}{\mbox{\boldmath $k$}_\beta}
\newcommand{\kc}{\mbox{\boldmath $k$}_c}
\newcommand{\kd}{\mbox{\boldmath $k$}_d}
\newcommand{\xxb}{\mbox{\boldmath $x$}}
\newcommand{\sk}{\sum_{k\mu}}
\newcommand{\ux}{\mbox{\boldmath $u$}}
\newcommand{\vx}{\mbox{\boldmath $v$}}
\newcommand{\ffx}{\mbox{\boldmath $f$}}
\newcommand{\gx}{\mbox{\boldmath $g$}}

\begin{document}

\maketitle
\begin{abstract}
Grangier, Roger and Aspect (GRA) performed a beam-splitter experiment to demonstrate the particle behaviour of light and  a Mach-Zehnder interferometer experiment to demonstrate the wave behaviour of light. The distinguishing feature of these experiments is the use of a gating system to produce near ideal single photon states. With the demonstration of both wave and particle behaviour (in two mutually exclusive experiments) they claim to have demonstrated the dual particle-wave behaviour of light and hence to have confirmed Bohr's principle of complementarity. The demonstration of the wave behaviour of light is not in dispute. But we want to demonstrate, contrary to the claims of GRA, that their beam-splitter experiment does not conclusively confirm  the particle behaviour of light, and hence does not confirm particle-wave duality, nor, more generally, does it confirm complementarity. Our demonstration consists of providing a detailed model based on the Causal Interpretation of Quantum Fields (CIEM), which does not involve the particle concept, of  GRA's which-path experiment.  We will also give a brief outline of a  CIEM model for the second, interference, GRA experiment.
\end{abstract}
\section{Introduction}
There are countless experiments which demonstrate the wave behaviour of light. Two typical experiments are the two-slit and Mach-Zehnder arrangements. That such experiments demonstrate the wave behaviour of light, even where the light is feeble\footnote{By feeble light we mean light of such low intensity that on average only one photon at a time is in the apparatus.} \cite{T09}, is not in dispute. What is questionable is the experimental evidence for the particle behaviour of light.

To avoid later misunderstanding of the essential point of this article, it is necessary for me to make clear that I use the term  `particle behaviour' to refer to the description prior to the final detected result but not to the character of the final detected result. This is a more restrictive usage than is usual in the literature where the term `particle behaviour' also encompasses the character of the final detected experimental result. I also use the term `particle behaviour' in two context dependent ways: In the context of Bohr's principle of complementarity I use the term `particle behaviour' to refer to the description of the experiment in terms of the complementary particle concept (understanding that according to Bohr the particle concept, along with other complementary concepts, is an abstraction to aid thought to which physical reality cannot be attached). In the context of the causal interpretation I take the term `particle behaviour' to be synonymous with `particle ontology'. Similar considerations apply to the term `wave behaviour', but the distinction here is not so crucial since a main point of this article is to demonstrate that a final detected result showing a particle character does not force a particle description or particle ontology prior to the final detected result.

More recent and interesting experiments concerning particle-wave duality and complementarity have been suggested and subsequently performed. Ghose {\it et al} \cite{GHOSE91} proposed an experiment involving tunneling between two closely spaced prisms which  has since been carried out by Mizobuchi {\it et al} \cite{MIZ92} (although the statistical results  of the experiment have been questioned by  \cite{UNNIK, GHOSE99, BRIDA04}). Later, Brida  {\it et al}  \cite{BRIDA04} realized  an experiment suggested by Ghose \cite{GHOSE99} in which tunneling at a twin prism arrangement is replaced by birefringence. Also of interest is Afshar's experiment \cite{AFSHAR04}. All of these experiments use light and aim to disprove or generalize\footnote{Brida {\it et al} view the observation of simultaneous particle and wave behaviour  as demonstrating a need to generalize complementarity in the sense of Wootters and Zurek \cite{WZ}and Greenberger and Yasin \cite{GY}. I have argued  that the generalization in fact completely contradicts complementarity and is the antithesis of Bohr's teachings \cite{KPW}. See section 6 for further discussion of this point.} complementarity (whereas GRA's aim was to confirm complementarity) by claiming to have demonstrated particle and wave behaviour in the same experiment. In all of these experiments, the final detection  result is attributed by the authors to which-path information and, therefore, to particle behaviour (according to the usual criteria accepted in the literature), but the experiments are so arranged that the light  undergoes a process (tunneling in the case of Mizobuchi {\it et al}'s experiment, birefringence in Brida {\it et al}'s  experiment, and interference in Afshar's experiment) which the authors claim necessarily represents wave behaviour. Hence, they claim to observe wave and particle behaviour in the same experiment. We do not agree with them for the same reasons that we do not agree with GRA's  claim to have proved complementarity, a claim we will argue against in this article. Generally, we take the view that complementarity is so imprecise that it can neither be proved nor disproved. We will elaborate further on this in the rest of the article with regard to the GRA experiments, but we will also briefly describe and comment further on Mizobuchi {\it et al}'s, Brida {\it et al}'s  and  Afshar's experiments in section \ref{CGBAS}. We have chosen to focus on the GRA experiments in this article because they were the first to introduce a gating system for producing genuine single photon states and because their experiments lend themselves to illustrating important features of CIEM. Further, the detailed treatment of this experiment serves as a model  that can be easily adapted to the later experiments, thereby providing arguments against the claims of observing simultaneous wave and particle behaviour in these experiments. The quantum eraser experiment of Kim {\it et al} \cite{KIM00}  is a variant of the Wheeler delayed-choice idea \cite{WHR78, K05}. The use of particle-wave duality and complementarity in this experiment seems to imply that a measurement performed in the present effects the outcome of an earlier measurement. This now raises the  further issue of the present effecting the past, which is surely unacceptable. We will also give a brief description and comment on this experiment in section \ref{CGBAS}. 

Experimental evidence for the particle behaviour of light is mainly of two forms: which-path experiments and the photoelectric effect (also the Compton effect). A closer look at each of these shows that neither unambiguously demonstrate particle behaviour. In the case of the photoelectric effect it is well known that a semiclassical description can be given in which the light is treated as a classical electromagnetic field and only the atom is treated quantum mechanically \cite{W26}. A weakness of this counter example  is that semiclassical radiation theory is known not to be fully consistent with experiment and fails  in those cases where light exhibits nonclassical properties (as in some experiments which involve second-order coherence). Further, it is not clear that a semiclassical model of the photoelectric effect can explain the experimental fact that the photon is absorbed in a time of the order of  $10^{-9}\;\mathrm{s}$ (\cite{VW76}, p. 10). Indeed, it was just this feature of the photoelectric effect that seemed to require that a photon be a localized particle prior to absorption, and is perhaps the reason why the photoelectric effect is commonly regarded as evidence for the particle behaviour of light. A more convincing argument against the photoelectric effect as evidence of particle behaviour is the provision of a fully quantum mechanical model of the photoelectric effect based on the causal interpretation of the electromagnetic field (CIEM) \cite{K85, K87, K94}. In CIEM, light is modeled as a real vector field; there are no photon particles\footnote{From here on we will use the term `photon' very loosly to refer to a quantum of energy which may or may not be spread out over large regions of space with a value  of $\hbar\omega$ for a Fock state or with a value an average around $\hbar\omega$ for a wave packet.}. The field has the property of being nonlocal, meaning that an interaction at one point in the field can change the field at points beyond $ct$. The CIEM model of the photoelectric effect is of the nonlocal absorption of a photon by a localized atom. The photon prior to absorption may be spread over large regions of space. The fact that the absorption is nonlocal explains the experimental result  that the absorption of  the photon takes place in a time of the order of $10^{-9}\;\mathrm{s}$. We are not forced to accept that the photon must be localized prior to absorption. We conclude that the photoelectric effect cannot be regarded as conclusive evidence for the particle behaviour of light.  We note that the  Compton effect, also commonly accepted as evidence for the particle behaviour of light, can also be modeled by CIEM (\cite{K94}, p. 343), so that this also cannot be taken as evidence for the particle behaviour of light. To be clear, we are not claiming that the final detected results of the photoelectric and the Compton effect do not have a particle character (they clearly do). What we claim is that  a particle description prior to the final result, whether from the perspective of complementarity or from the perspective of an ontology, is not forced upon us. This is because the particle character of the experimental results can be explained in terms of a wave model.

Let us now turn to which-path experiments. In a typical which-path experiment light has a choice of two paths. Determining 
which-path the light actually took is considered as proof of particle behaviour. As Bohr showed in response to Einstein's famous 
which-path two-slit experiment,  if the path is determined with certainty, interference is lost \cite{BR59A}. Consider a which-path two-slit experiment in which we determine the path by closing one of the holes (obviously losing interference). Although crude, it is conceptually equivalent to Einstein's experiment. The point is, that even when we close the hole and are certain which-path the light took, this does not rule out a wave model. This argument  holds even in more refined which-path two-slit experiments. We may conclude that in such experiments the which-path criteria for particle behaviour is somewhat arbitrary.

There is an aspect of the two-slit experiment that seems to be universally overlooked and that we wish to draw attention to. Einstein's aim in his which-path two-slit experiment was to obtain the path of an individual photon and still retain an interference pattern, thereby  experimentally detecting particle and wave behaviour in the same experiment\footnote{Actually, Einstein considered Bohr's principle of  complementarity and quantum mechanics to be synonymous. By experimentally contradicting complementarity Einstein wanted to  demonstrate that quantum mechanics is incomplete (\cite{JAM74}, p. 127). We have argued elsewhere  that Bohr's principle of complementarity and quantum mechanics are not synonymous (\cite{K05}, p. 299).}. This is contrary to Bohr's principle of complementarity which requires mutually exclusive experimental arrangements for complementary concepts \cite{BR59A, JAM74, BR28}. As we have said, Bohr was able to show that a certain determination of the photon path would destroy the interference pattern.  Bohr's response was almost universally accepted and complementarity was saved. But consider this: Forget path determination and consider a two-slit experiment in which an interference pattern is formed. This interference pattern is built up of a large number of individual  photoelectric detections (or some similar process in a photographic emulsion). If the photoelectric effect is accepted as evidence of the particle behaviour of light, then is not particle and wave behaviour observed in the same experiment?

We now turn to another which-path experiment which uses a beam-splitter. This will be our main focus in this article because we consider GRA's version of this experiment, which uses an atomic cascade and a gating system to produce a near ideal single photon state, as perhaps the best experimental attempt to demonstrate the particle behaviour of light \cite{G86, AG86}. In a wave model, light is split into two beams at the beam-splitter. In a particle model, each photon must choose one and only  one path. Thus, using feeble light (one photon at a time) a particle model predicts perfect anticoincidence, whereas some coincidences are expected in a wave model. GRA therefore took perfect anticoincidence as the signature of particle behaviour. GRA quantified this feature in terms of the degree of second-order coherence. Semiclassical radiation theory predicts $g^{(2)}\geq 1$. As we shall see, quantum mechanical coherent or chaotic states give results in the classical regime.  This is to be expected, as neither chaotic nor coherent light exhibits nonclassical behaviour. For number states on the other hand,  perfect anticoincidence is expected, so that $g^{(2)}=0$.    

Photoelectric detectors are placed in each output arm of the beam-splitter. For a detection to take place there must be enough energy to ionize an atom in the detector. For classical light, and quantum mechanical chaotic or coherent light,  there is always some probability that more than one photon is present  after the beam-splitter however feeble the light, and this entails the possibility of coincidences. But, for a single photon state there is enough energy to ionize only a single atom in one and only one output arm of the beam-splitter, so that perfect anticoincidence is predicted.

The novelty of the GRA experiments is the use of an atomic cascade and a gating system, which we describe below, in order to produce near ideal single photon states. Their results gave a value of $g^{(2)}$ much less than $1$ and confirmed the expected anticoincidence. GRA interpreted their results to be a conclusive demonstration of the particle behaviour of light. 

But, underlying the assertion that anticoincidence is a signature for particle behaviour is the assumption that the  photoelectric detection process (or any other atomic absorption process) is local. This implies that the photon is a localized particle before absorption by  the detecting atom. But, we saw above that the quantum theory does not rule out nonlocal absorption in the photoelectric effect (nor, more generally,  in any atomic absorption process). In fact, no model of light as photon particles that is consistent with the quantum theory has ever been developed\footnote{Ghose {\it et al} have developed a particle interpretation of bosons \cite{GHOSE93, GHOSE96}, including the photon \cite{GHOSE01}, based on the Kemmer-Duffin formalism \cite{KEMMER39}. It is to be emphasized that this formalism, which allows an interpretation of bosons as particles, applies in the approximation that the energies are below the threshold for pair production. We maintain that the full theory does not allow a particle ontology. Since the particle ontology of the approximation stands in contradiction to the ontology of the full field theory  (since particle and wave concepts are mutually exclusive), we maintain that the particle ontology of the approximate theory cannot have physical significance (Ghose {\it et al} do not address this issue). A further point is this: As Ghose himself points out, reference (\cite{GHOSE96}, p. 1448), for the boson particle interpretation to be consistent negative energy solutions must be interpreted as antiparticles moving backwards in time. In this case, an EPR correlated particle-antiparticle pair would exhibit the pathological feature of a  nonlocal connection between the present and the past (we note that this particular criticism does not apply to the electromagnetic field). For more details on this and related approaches see reference \cite{WS2005}.}. On the other hand, CIEM  models light as a nonlocal field. Atomic absorption  processes, including the photoelectric effect, are modeled  as the nonlocal absorption of a photon. CIEM has been shown to be fully consistent with the quantum theory \cite{K94}. Our main purpose in this article is to provide a model that explains perfect anticoincidence that does not  treat photons as particles. By showing that anticoincidence experiments do not rule out a wave model we prove that GRA's experiment cannot be viewed as conclusive evidence for particle behaviour of light. 

The wave behaviour of light has been confirmed a countless number of times for  chaotic or coherent sources. Following Einstein's 1905 explanation of the photoelectric effect \cite{E1905} in which the idea of photon particles was first invoked,  the question was raised as to whether or not, in very low intensity experiments, single photons alone in the apparatus can produce interference. Numerous experiments using feeble light followed \cite{T09}. With a few exceptions the conclusion was reached that single photons can interfere with themselves. In  such experiments the energy flux $\cal{E}$ is calculated and the number of photons per unit area per unit time is calculated using ${\cal E}/ \hbar \omega$. $\cal{E}$ is reduced to such low levels that it is more probable than not that only one photon is present in the apparatus at any one time. However, the probability that more than one photon is present remains, so that the single photon nature of these experiments can be questioned. By building a Mach-Zehnder interferometer around their which-path apparatus GRA were able to confirm that the near ideal single photon state produced the expected interference. Although no surprise, GRA's experiment is perhaps the first experiment to confirm the interference of single photons. The wave nature of light is not disputed and it is obvious how  in CIEM interference is obtained given that light is modeled as a field (always). We will nevertheless outline the CIEM treatment of the Mach-Zehnder interferometer given in detail in reference \cite{K05}.

CIEM is a hidden variable theory. There is a large  literature on hidden variable theories and we direct the interested reader to the three articles cited in  reference \cite{HVTHR}. Two of these, one old one new, are surveys of hidden variable theories and include a comprehensive list of references. We also refer the reader to two interesting Ph.D thesis in the area of hidden variable theories
\cite{SC2005, WS2005}.  

In the next sections we describe GRA's two experiments focusing on theoretical derivations, and then go on to give the CIEM model of these experiments, focusing on the which-path experiment.

\section{The GRA experiments}
The following description of the GRA experiments is based mainly on reference \cite{G86}. The experiments use the radiative cascade of calcium $4p^2\;^1S_0\rightarrow4s4p\;^1P_1\rightarrow4s^2\;^1S_0$ described in reference \cite{AGR81}. The first cascade to the intermediate state yields a photon $\nu_1$ of wavelength $551.3\;\mathrm{nm}$. The intermediate state, with lifetime $\tau=4.7\;\mathrm{ns}$, decays according to the usual atomic decay law for the lifetime of a state (\cite{BJ89}, p. 538):
\begin{equation}
P(t)=1-e^{-t/\tau}, \label{DL}
\end{equation}
where $P(t)$ is the probability of decay in time $t$. The second cascade photon $\nu_2$ has  wavelength $422.7\;\mathrm{nm}$. The $\nu_2$ photon, according to the decay law, is emitted with near certainty within the time  $\omega=2\tau=9.8\;\mathrm{ns}$ of emission of the first $\nu_1$ photon. The number of $\nu_1$ photons per second, $N_1$, is counted by photomultiplier $PM_1$, and each $\nu_1$ photon triggers a gate of duration $\omega$. Because the probability of decay within gate $\omega$ is high, there is a high probability that the $\nu_2$ partner of $\nu_1$ enters the beam-splitter. For low count rates we can be nearly certain that there is only one $\nu_2$ photon in the beam-splitter arrangement within the gate time $\omega$. In this way a near ideal single photon state is produced.

\begin{figure}[h]
\unitlength=1in
\hspace*{.7in}\includegraphics[width=5in,height=2.2in]{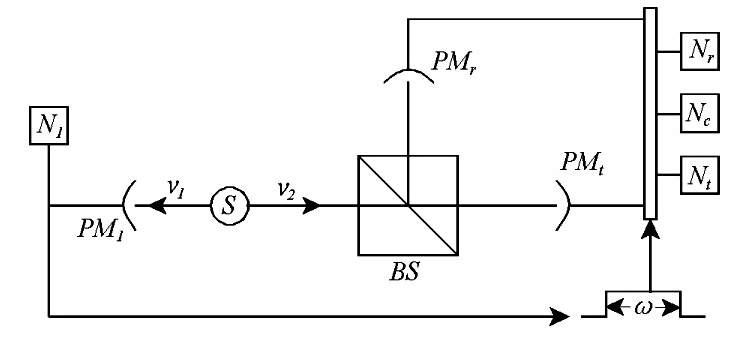}
\caption{GRA's which-path experiment}
\label{GRA1}
\end{figure}

\section{GRA's which-path experiment}
Refer to figure \ref{GRA1}. The photomultipliers $PM_t$ and $PM_r$ count the number of transmitted and reflected $\nu_2$ photons per second, and photomultiplier $PM_c$ counts the number of coincidences per second. These count rates are given by $N_t$, $N_r$ and $N_c$ respectively. The counts are taken over a large number of gates with a total run time $T$ of about 5 hours. The probabilities for single and coincidence counts are given by
\begin{equation}
p_t=\frac{N_t}{N_1},\;\;\;\;\;\;\;\;p_r=\frac{N_r}{N_1},\;\;\;\;\;\;\;\;p_c=\frac{N_c}{N_1}.
\end{equation}
The classical and quantum mechanical predictions for the coincidence counts are very different.
In their experiment, GRA measured the quantity $\alpha$, which they defined as \cite{G86}
\begin{equation}
\alpha=\frac{\mbox{\it \small COINCIDENCE PROBABILITY}}{\mbox{\it \small ACCIDENTAL COINCIDENCE PROBABILITY}}=\frac{p_c}{p_t p_r}=\frac{N_1N_c}{N_t N_r}. \label{alpha}
\end{equation}
Both classically and quantum mechanically, the quantity $\alpha$ is a special case of the degree of second-order coherence. Classically, $g^{(2)}_c$ is defined by (\cite{L73}, p. 111)
\begin{equation}
g^{(2)}_c(\rvect_1t_1,\rvect_2 t_2; \rvect_2 t_2, \rvect_1t_1)=\frac{\langle \E^*(\rvect_1t_1)\E^*(\rvect_2t_2)\E(\rvect_2t_2)\E(\rvect_1t_1)\rangle}{\langle|\E(\rvect_1t_1)|^2 \rangle \langle|\E(\rvect_2t_2)|^2\rangle},
\end{equation}
where $\E$ is the electric field vector. For $\rvect_1=\rvect_2$ and $t_1=t_2$,  $g^{(2)}_c$ reduces to
\begin{equation}
g^{(2)}_c=\frac{\langle(\E^*\E)^2\rangle}{\langle\E^*\E\rangle \langle\E^*\E\rangle}=\frac{\langle I^2\rangle}{\langle I \rangle^2}, \label{DSOC}
\end{equation}
where $I$ is the intensity. We will see in the next subsection that $\alpha=g^{(2)}_c$.  Similar definitions apply in quantum mechanics (\cite{L73}, p. 219):
\begin{equation}
g^{(2)}(\rvect_1t_1,\rvect_2 t_2; \rvect_2 t_2, \rvect_1t_1)=\frac{\langle \Eh^-(\rvect_1t_1)
\Eh^-(\rvect_2t_2)\Eh^+(\rvect_2t_2)\Eh^+(\rvect_1t_1)\rangle}{\langle \Eh^-(\rvect_1t_1)\Eh^+(\rvect_1t_1)  \rangle \langle\Eh^-(\rvect_2t_2)\Eh^+(\rvect_2t_2) 
\rangle}, \label{G2}
\end{equation}
where the $\Eh$'s are quantum mechanical operators defined by  
\begin{equation}
\Eh^+(\rvect t)=\frac{i}{\V}\sk\sqrt{\frac{\hbar k c}{2}}\ek\akk e^{i(\kb.\xb-\omega_k t)}, \;\;\;
\Eh^-(\rvect t) =-\frac{i}{\V}\sk\sqrt{\frac{\hbar k c}{2}}\ek\akkd  e^{-i(\kb.\xb-\omega_k t)}. \label{EPEM}
\end{equation}
By substituting eq. (\ref{EPEM}) into eq. (\ref{G2}) with $\rvect_1=\rvect_2$ and $t_1=t_2$   and considering only a single mode and a single polarization direction, eq. (\ref{G2}) reduces to
\begin{equation}
g^{(2)}=\frac{\langle a_{2}^{\dagger}a_2 a_{1}^{\dagger}a_1\rangle}{\langle a_{1}^{\dagger}a_1\rangle \langle a_{2}^{\dagger}a_2\rangle}.\label{G2qm}
\end{equation}
For a single mode and single polarization direction, the quantum mechanical operator for the magnitude of the intensity  (\cite{L73}, p. 184; \cite{K05}, p. 304) reduces to
\begin{equation}
\hat{I}_1=\frac{\hbar k c^2 }{V}a^{\dagger}_1a_1.
\end{equation}
Multiplying the numerator and the denominator of eq. (\ref{G2qm}) by $(\hbar k c^2/V)^2$, we can write $g^{(2)}$ in terms of the expectation value of the intensity operator:
\begin{equation}
g^{(2)}=\frac{\langle I_1 I_2\rangle}{\langle I_1 \rangle \langle I_2\rangle}.\label{G2qmI}
\end{equation}
Again, we will see in the next subsection that this is equivalent to GRA's $\alpha$.

In the following subsections we calculate the classical prediction for $g^{(2)}$ using semiclassical radiation theory and compare this with the quantum mechanical predictions for $g^{(2)}$ for a number state, a coherent state, and a chaotic state.

\subsection{$g_c^{(2)}$ for a classical field}
We now calculate the classical prediction for the various probabilities. The intensity of the $n^{th}$ gate  is given by the time average of the instantaneous intensity $I(t)$:
\begin{equation}
i_n=\frac{1}{\omega}\int^{t_n+\omega}_{t_n} I(t)\;dt.
\end{equation} 
Although the electromagnetic field is treated classically, the photoelectric detection is treated quantum mechanically. This  semiclassical radiation theory gives the probability for a detection as proportional to the intensity and to time (\cite{L73}, p. 183 and p. 185; \cite{M76} p. 31 and p. 40) (as is the case quantum mechanically). The probabilities for singles counts during the $n^{th}$ gate are, therefore,
\begin{equation}
p_{tn}=\alpha_t i_n\omega,\;\;\;\;\;\;\;\;\;\;\;p_{rn}=\alpha_r i_n\omega,
\end{equation}
where  $\alpha_t$ and $\alpha_r$ are the global detection efficiencies.  The intensity averaged over all the gates is
\begin{equation}
\langle i_n \rangle =\frac{1}{N_1 T}\sum^{N_1 T}_{n=1} i_n,
\end{equation}
where $N_1T$ is the total number of counts in $PM_1$, which is equal to the total number of gates. So, the overall probability for singles counts becomes
\begin{equation}
p_t=\alpha_t\omega\langle i_n\rangle,\;\;\;\;\;\;\;\;\;\;p_r=\alpha_r\omega\langle i_n \rangle.\label{PtPr}
\end{equation}
During a single gate, the probability of a detection in one arm is statistically independent of detection in the other arm. Therefore, the probability of  a coincidence count during a single gate is given as the product of the probabilities of detection in each arm:
\begin{equation}
p_{cn}=\alpha_t\alpha_r\omega^2 i_n^2.
\end{equation}
The probability of a coincidence count averaged over all the gates becomes
\begin{equation}
p_c=\alpha_t\alpha_r\omega^2\langle i_n^2\rangle.\label{Pc}
\end{equation}
If the coincidences are purely accidental, then the probabilities $p_t$ and $p_r$  over the ensemble of all gates are statistically independent, so that the accidental coincidence probability is given by the product
\begin{equation}
p_t p_r=\alpha_t\alpha_r\omega^2\langle i_n\rangle^2. \label{Ptr}
\end{equation}
This represents the minimum classical probability of coincidence. These averages satisfy the inequality (\cite{BS73}, p. 185, inequality no. 4)
\begin{equation}
\langle i_n^2 \rangle \geq \langle i_n \rangle^2, \label{PcPtPr}
\end{equation}
from which it follows, by using eq.'s (\ref{Pc}) and (\ref{Ptr}), that
\begin{equation}
p_c \geq p_t p_r.
\end{equation}
In terms of $\alpha$, eq. (\ref{alpha}), we can also write the inequality (\ref{PcPtPr}) as
\begin{equation}
\alpha\geq 1. \label{G2c}
\end{equation}
Substituting eqs. (\ref{Pc}) and (\ref{Ptr}) into eq. (\ref{alpha}) gives
\begin{equation}
\alpha=\frac{\langle i_n^2 \rangle}{\langle i_n \rangle^2}, \end{equation}
which is equal to the classical second-order coherence function $g_c^{(2)}$ given in eq. (\ref{DSOC}).

\subsection{Quantum mechanical $g^{(2)}$ for a number state, a coherent state and a chaotic state}
In quantum mechanics, the same reasoning as for the classical case leads to the same expressions for the probabilities $p_t$, $p_r$ and $p_c$, and for $\alpha$. The difference  is that  the classical averages of the intensities are replaced by quantum mechanical expectation values of the intensity operator. Thus
\begin{equation}
\alpha=\frac{p_c}{p_t p_r}=\frac{\alpha_t\alpha_r\omega^2\langle I_{\alpha} I_{\beta} \rangle}{\alpha_t\omega\langle I_{\alpha}\rangle\alpha_r\omega\langle I_{\beta}\rangle}=\frac{\langle I_{\alpha} I_{\beta} \rangle}{\langle I_{\alpha}\rangle\langle I_{\beta}\rangle}=\frac{\langle b^{\dag}_{\alpha} b_{\alpha} b^{\dag}_{\beta} b_{\beta} \rangle}{\langle b^{\dag}_{\alpha} b_{\alpha}\rangle\langle b^{\dag}_{\beta} b_{\beta}\rangle}. \label{alphaqm}
\end{equation}
The subscripts $\alpha$ and $\beta$ refer to the horizontal and vertical beams that emerge after the first beam-splitter. We see that $\alpha$ is equal to $g^{(2)}$, eq. (\ref{G2qm}) or eq. (\ref{G2qmI}), in the quantum case also. 

To calculate $g^{(2)}$ we first consider the theoretical treatment of a single beam-splitter. By now a two input approach to the 
beam-splitter is almost universally accepted even when one of the inputs is the vacuum\footnote{In passing, we mention  that Caves \cite{C80} uses a two input approach in connection with the search for gravitational waves  using a Michelson interferometer. He suggests, as one of two possible explanations, that vacuum fluctuations due to a vacuum input are responsible for the `standard quantum limit' which places a limit on the accuracy of any measurement of the position of a free mass.}  (e.g. \cite{OHM87}), but some  workers still use a single input (\cite{L73}, p. 222\footnote{Here the beam-splitter is described as part of the Hanbury-Brown and Twiss experiment.}; \cite{SZ97}, p. 494\footnote{Here the beam-splitter is used as part of an atomic interferometer.}). The two input approach leads to an elegant mathematical description of the action of a beam-splitter in terms of a unitary $2\times 2$ transformation matrix which has the form of a rotation matrix \cite{CST}. Here we will use a use a single input approach since this greatly simplifies the mathematical treatment of the GRA experiments in terms of CIEM, and since it gives the same results as the two input approach for the quantities we are interested in (expectation values of the number operator, coincidence counts, and interference terms). Further, both approaches lead to essentially the same physical model of the GRA experiments in terms of CIEM.

\begin{figure}[h]
\unitlength=1in
\hspace*{1.5in}\includegraphics[width=3in,height=1.8in]{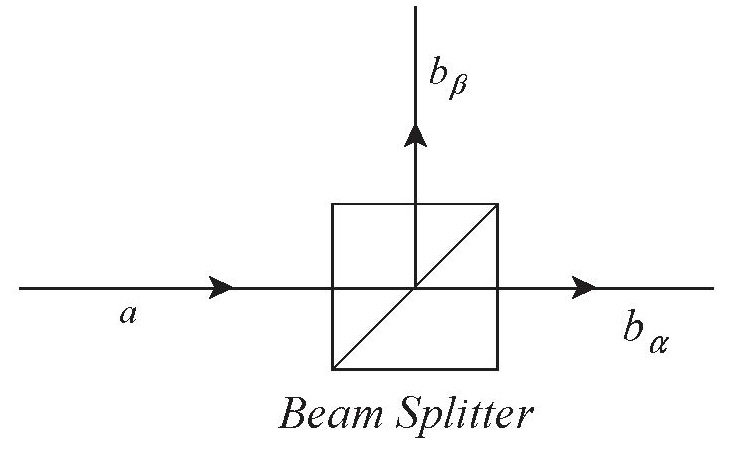}
\caption{Input and output destruction operators.}
\label{INOUTBS}
\end{figure}

The single input and two output annihilation and creation operators are related as follows:
\begin{equation} 
a=t^{*}_{{\alpha}{\alpha}}b_{\alpha}+r^{*}_{{\alpha}{\beta}}b_{\beta},\;\;\;\;\;\;\;\;\;\;\;a^{\dagger}_{\alpha}=t_{{\alpha}{\alpha}}b^{\dagger}_{\alpha}+r_{{\alpha}{\beta}}b^{\dagger}_{\beta}. \label{cao}
\end{equation}
The $b$'s satisfy the usual commutation relation $[b_{\alpha}, b^{\dagger}_{\alpha}]=[b_{\beta}, b^{\dagger}_{\beta}]=1$ while any combination of $b_{\alpha}$ and $b_{\beta}$ or their conjugates commute. To preserve the commutator $[a, a^{\dagger}]=1$, we must have
\begin{equation} 
|t_{{\alpha}{\alpha}}|^2+|r_{{\alpha}{\beta}}|^2=t^2+r^2=1, \label{TR}
\end{equation}
with $|t_{{\alpha}{\alpha}}|^2=t^2$ and $|r_{{\alpha}{\beta}}|^2=r^2$. Using eq.'s (\ref{cao}) and (\ref{TR}) we may proceed to calculate $g^{(2)}$ for various quantum states. We begin with the number state $| n \rangle$, 
\begin{equation}
| n \rangle=\frac{(a^{\dagger}_{\alpha})^n}{(n!)^{\frac{1}{2}}}| 0 \rangle=\frac{(t_{{\alpha}{\alpha}}b^{\dagger}_{\alpha}+r_{{\alpha}{\beta}}b^{\dagger}_{\beta})^n}{(n!)^{\frac{1}{2}}}| 0 \rangle.
\end{equation}
Use of  the binomial theorem to expand the brackets gives
\begin{eqnarray}
| n \rangle&=&\frac{1}{(n!)^{\frac{1}{2}}}\left[ \left( \begin{array}{c}n\\0 \end{array}\right) (t_{{\alpha}{\alpha}}b^{\dagger}_{\alpha})^n +\left( \begin{array}{c}n\\1 \end{array}\right) (t_{{\alpha}{\alpha}} b^{\dagger}_{\alpha})^{(n-1)} (r_{{\alpha}{\beta}}b^{\dagger}_{\beta})^1\right. \nonumber\\
& &+ \left( \begin{array}{c}n\\2 \end{array}\right) (t_{{\alpha}{\alpha}} b^{\dagger}_{\alpha})^{(n-2)} (r_{{\alpha}{\beta}}b^{\dagger}_{\beta})^2+......
+\left( \begin{array}{c}n\\n-1 \end{array}\right) (t_{{\alpha}{\alpha}} b^{\dagger}_{\alpha})^1 (r_{{\alpha}{\beta}}b^{\dagger}_{\beta})^{(n-1)}\nonumber\\
&&\left. +\left( \begin{array}{c}n\\n \end{array}\right) (r_{{\alpha}{\beta}} b^{\dagger}_{\beta})^n \right]| 0 \rangle. 
\end{eqnarray}
With this expression for  $|n\rangle$ we can evaluate the expectation value for the number of photons in the horizontal arm, $\langle n| b^{\dagger}_{\alpha} b_{\alpha} |n\rangle$,  by multiplying out the brackets, noting that cross-terms are zero, and evaluating the action of the number operator on the various number states. After a number of rearrangement steps we arrive at
\begin{eqnarray}
\langle b^{\dagger}_{\alpha} b_{\alpha}\rangle=\langle n| b^{\dagger}_{\alpha} b_{\alpha} |n\rangle&=&nt^2\left[ t^{2(n-1)}+
t^{2(n-2)}r^2 \frac{(n-1)!}{(n-2)!}+ t^{2(n-3)}r^4\frac{(n-1)!}{(n-3)!2!}\right.\nonumber\\
&&\left.+t^{2(n-4)}r^6\frac{(n-1)!}{(n-4)!3!}+......+r^{2(n-1)}\right].
\end{eqnarray}
We recognize the series in the square brackets as the binomial expansion for $(t^2+r^2)^{n-1}=1$, and we get
\begin{equation}
\langle b^{\dagger}_{\alpha} b_{\alpha} \rangle=nt^2. \label{bhbh}
\end{equation}
By the same procedure as above we also get the expectation value for the number of photons in the vertical beam,
\begin{equation}
\langle b^{\dagger}_{\beta} b_{\beta}\rangle=\langle n| b^{\dagger}_{\beta} b_{\beta} |n\rangle=nr^2, \label{bVbV}
\end{equation}
and the expectation value for the number of coincidences,
\begin{equation}
\langle b^{\dagger}_{\alpha} b_{\alpha}b^{\dagger}_{\beta} b_{\beta} \rangle=\langle n| b^{\dagger}_{\alpha} b_{\alpha}b^{\dagger}_{\beta} b_{\beta} |n\rangle=n(n-1)r^2t^2.\label{bHbHbVbV}
\end{equation}
Substituting the above expectation values into eq. (\ref{G2qm}) gives the second-order coherence function for a number state,
\begin{equation}
g^{(2)}=\frac{n(n-1)r^2t^2}{nt^2nr^2}=\frac{(n-1)}{n},\;\;\;\;\;\;n\geq 2.
\end{equation}
For $n=0,1$ $g^{(2)}$=0. We see that a single photon input shows perfect anticorrelation, contrary to the classical result for $g^{(2)}_c$, eq. (\ref{G2c}).
Next we consider the coherent state
\begin{equation}
|\alpha\rangle=e^{-|\alpha|^2/2}\sum_n \frac{\alpha^n}{(n!)^{\frac{1}{2}}}|n\rangle.
\end{equation}
The expectation value in the horizontal arm is
\begin{equation}
\langle b^{\dagger}_{\alpha} b_{\alpha}\rangle=\langle\alpha| b^{\dagger}_{\alpha}b_{\alpha}|\alpha\rangle=e^{-|\alpha|^2}\sum_{n=0} \frac{|\alpha|^{2n}}{n!}\langle n| b^{\dagger}_{\alpha}b_{\alpha}|n\rangle +e^{-|\alpha|^2}\sum_{n'}\sum_{\stackrel{\mbox{\scriptsize$n$}}{\!\!\!\!\!\!\!\!\!\!\!n\neq n'}}
\frac{(\alpha^{*})^{n'}}{(n'!)^{\frac{1}{2}}}\frac{\alpha^{n}}{(n!)^{\frac{1}{2}}}\langle n'| b^{\dagger}_{\alpha}b_{\alpha}|n\rangle.
\end{equation}
The second term consisting  of cross terms is zero. After substituting eq. (\ref{bhbh}) into the above, we get
\begin{equation}
\langle b^{\dagger}_{\alpha}b_{\alpha}\rangle=t^2 e^{-|\alpha|^2}\sum_{n=0} \frac{|\alpha|^{2n}}{n!}n=t^2 e^{-|\alpha|^2}|\alpha|^2\sum_{n=0} \frac{|\alpha|^{2n}}{n!}=t^2 e^{-|\alpha|^2}|\alpha|^2e^{|\alpha|^2}
=t^2 |\alpha|^2.
\end{equation}
In a similar way,  we calculate the expectation value of the number operator in the vertical beam to be
\begin{equation}
\langle b^{\dagger}_{\beta} b_{\beta} \rangle=\langle\alpha| b^{\dagger}_{\beta}b_{\beta}|\alpha\rangle=r^2 |\alpha|^2,
\end{equation}
and the expectation value for coincidence counts to be
\begin{equation}
\langle b^{\dagger}_{\alpha} b_{\alpha}b^{\dagger}_{\beta} b_{\beta} \rangle=\langle\alpha| b^{\dagger}_{\alpha}b_{\alpha}b^{\dagger}_{\beta}b_{\beta}|\alpha\rangle=t^2r^2 |\alpha|^4.
\end{equation}
Substituting the above expectation values into eq. (\ref{G2qm}) gives the second-order coherence function for a coherent state as
\begin{equation}
g^{(2)}=\frac{t^2r^2 |\alpha|^4}{t^2 |\alpha|^2r^2 |\alpha|^2}=1.
\end{equation}
This corresponds to the minimum classical value for $g^{(2)}$ so that measurement of the degree of second order coherence cannot distinguish between classical and coherent light. 

Lastly, we consider chaotic light. In quantum mechanics, chaotic light is a mixture of number states and is represented by the density operator (\cite{L73}, p. 158)
\begin{equation}
\rho=\sum_n P_n|n\rangle \langle n|.
\end{equation} 
For light in thermal equilibrium, let $P_n$ be the probability of occurance of a number state $|n\rangle$ with energy $E_n=n\hbar\omega$. The probability $P_n$ is given by the Boltzmann distribution law applied to discrete quantum states (\cite{L73}, p. 8),
\begin{equation}
P_n=\frac{e^{-n\hbar\omega/kT}}{\sum^{\infty}_{n=0}e^{-n\hbar\omega/kT}}=(1-e^{-\hbar\omega/kT})\sum_n e^{-n\hbar\omega/kT},
\end{equation}
where $k$ is Boltzmann's constant, and  $T$ is the temperature in degrees Kelvin. The expectation value of the horizontal beam number operator is
\begin{eqnarray}
\langle b^{\dagger}_{\alpha} b_{\alpha} \rangle&=&{\mathrm Tr}(\rho b^{\dag}_{\alpha} b_{\alpha})=\sum_{n'}\langle n' |\rho b^{\dagger}_{\alpha} b_{\alpha}| n' \rangle=\sum_{n'} \sum_n(1-U)U^n \langle n'| n \rangle\langle n|b^{\dagger}_{\alpha} b_{\alpha}| n' \rangle \nonumber\\
&= &(1-U)\sum_n U^n \langle n| b^{\dagger}_{\alpha} b_{\alpha}| n \rangle, 
\end{eqnarray}
with $U=\exp(-\hbar\omega/kT)$. Substituting the expectation value (\ref{bhbh}), and rearranging gives
\begin{equation}
\langle b^{\dagger}_{\alpha} b_{\alpha}\rangle=t^2 \frac{U}{1-U}.
\end{equation}
Using the other expectation values for the number state as above, we easily get the results
\begin{equation}
\langle  b^{\dagger}_{\beta} b_{\beta} \rangle=r^2\frac{U}{1-U},\;\;\;\;\;\;\langle b^{\dagger}_{\alpha} b_{\alpha}b^{\dagger}_{\beta} b_{\beta} \rangle=t^2r^2 \frac{2U^2}{(1-U)^2}.
\end{equation}
Substituting the above into eq. (\ref{G2qm}) gives the degree of second-order coherence for a chaotic state
\begin{equation}
g^{(2)}=2
\end{equation}
Like the result with the coherent state this value lies in the classical range. 

\subsection{Comparison of theoretical and experimental results}
GRA's arrangement, figure \ref{GRA1}, gives the degree of second-order coherence $g^{(2)}$ directly by measurement of $N_t$, $N_r$ and $N_c$ and use of eq. (\ref{alpha}). A value of $g^{(2)}\geq 1$  would agree with classical mechanics while a zero value would confirm quantum mechanics. In practice,  experimental error  prevents an exact zero value. Therefore, before comparing experimental and theoretical results, we first derive, following GRA \cite{G86}, a practical quantum mechanical prediction. 

\begin{figure}[h]
\unitlength=1in
\hspace*{.8in}\includegraphics[width=4in,height=2.5in]{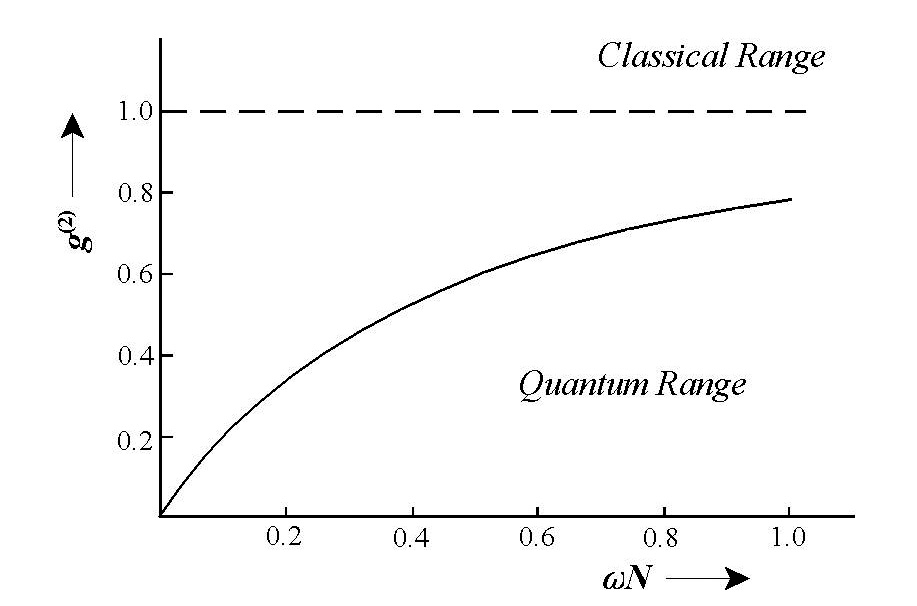}
\caption{Plot of the function $g^{(2)}(N\omega)$ with $f(w)=0.9$. 
\label{G2qmexp}}
\end{figure}

Let $N$ be the number of decays per second in the window of photomultiplier $PM_1$ of efficiency $\epsilon_1$. Then, $N_1=\epsilon_1 N$ is the number of $\nu_1$ photons detected per second by $PM_1$. From the atomic decay law (\ref{DL}), the probability $P_2$ of a $\nu_2$ photon partner of  a $\nu_1$ photon entering the beam-splitter during a gate $\omega$ triggered by $\nu_1$  is $1-\exp(-\omega/\tau)$. Because  of the angular correlation between $\nu_1$ and $\nu_2$, the probability $P_2$ is increased by a factor $a$ slightly greater than  $1$ \cite{F73}. This probability is denoted by  $f(\omega)=a[1-\exp(-\omega/\tau)]$, and is a number close to $1$ in GRA's experiment.The probability $P_2$ is also increased by accidental $\nu_2$'s. These are $\nu_2$ photons that enter the beam-splitter whose $\nu_1$ partners do not trigger a gate $\omega$. Once a $\nu_1$ photon has triggered a gate, the $\nu_1$ photons resulting from  $N\omega$ decays during the gate $\omega$ cannot trigger another decay. Hence, their $N\omega$ $\nu_2$ partners are the accidental $\nu_2$ photons. Since $N\omega$ is the number of accidental $\nu_2$'s entering the beam-splitter during gate $\omega$, then $N_1N\omega$ is the number of accidental $\nu_2$'s entering the beam-splitter per second. The probability of an accidental $\nu_2$ photon entering the beam-splitter is therefore $N_1N\omega/N_1=N\omega$. Thus,
\begin{equation}
P_2=f(\omega)+N\omega=\frac{N_2}{N_1},
\end{equation}
where
\begin{equation}
N_2=N_1[f(\omega)+N\omega]
\end{equation}
is the number of $\nu_2$ photons that enter the beam-splitter per second. Now, define $\epsilon_t$ and $\epsilon_r$ to be the efficiencies of $PM_t$ and $PM_r$, respectively. These efficiencies include the reflection and transmission coefficients, the collection solid angle, and the detector efficiency. The number $N_t$ of $v_2$ photons transmitted is $N_t=\epsilon_tN_2$, while the number reflected is $\epsilon_r N_2$. Then, the probabilities of detecting a transmitted $v_2$ photon in $PM_t$ and a reflected $v_2$ in $PM_r$ are
\begin{equation}
p_t=\frac{N_t}{N_1}=\frac{\epsilon_tN_2}{N_1} =\epsilon_t\left[f(\omega)+N\omega\right], \;\;\;\;\;\;\;\;\;\;\\
p_r=\frac{N_r}{N_1}=\frac{\epsilon_r N_2}{N_1} =\epsilon_r\left[ f(\omega)+N\omega\right].
\end{equation} 
Since $p_t$ and $p_r$ are statistically independent classically, the probability of a coincidence count becomes
\begin{equation}
p_c=p_tp_r=\epsilon_t\epsilon_r\left[ f(\omega)+N\omega\right]^2
=\epsilon_t\epsilon_r\left[ f(\omega)^2+2N\omega f(\omega) +N^2\omega^2\right].\label{PC}
\end{equation}
The term $f(\omega)^2$ suggests a repeated detection of the same photon. Since this is not possible,  $f(\omega)^2$ is set equal to zero. Thus, substituting $f(\omega)^2=0$ into eq. (\ref{PC}) gives the quantum mechanical  experimental expression for $p_c$. Substituting $p_t$, $p_r$ and $p_c$ into eq. (\ref{alpha}) gives:
\begin{equation}
g^{(2)}(N\omega)=\frac{2N\omega f(\omega) + N^2\omega^2}{[f(\omega)+N\omega]^2}.
\end{equation} 
A plot of this function is given in figure \ref{G2qmexp}.  It is noticeable that as the erroneous $N\omega$ $\nu_2$ photon count increases compared to $f(\omega)$ the value of $g^{(2)}$ approaches the classical minimum value. GRA's experimental results closely agree with the plot of figure \ref{G2qmexp}, and therefore confirm the quantum mechanical anticorrelation of the two beams. 

\section{GRA's Interference Experiment}

\begin{figure}[h]
\unitlength=1in
\hspace*{1in} \includegraphics[width=4in,height=2.2in]{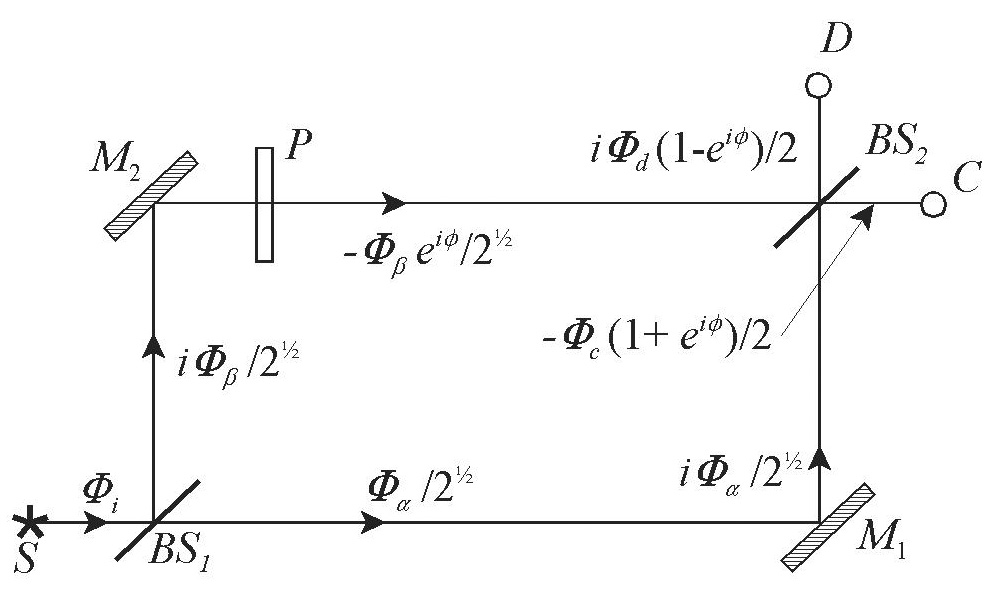}
\caption{GRA's interference experiment. The experiment uses the same novel gating
system (not shown) to produce a near ideal single photon state as in GRA's which-path 
experiment.}
\label{GRAmz}
\end{figure}

In the second interference experiment, GRA built a Mach-Zehnder interferometer around the first beam-splitter as shown in figure \ref{GRAmz}. Quantum mechanics predicts that each beam is oppositely modulated and that the fringe visibility of each beam as a function of path difference (or of a phase shift produced by a phase shifter) is $1$. In the experiment, interference fringes with visibility greater than 98\%  were observed.  Although the interference is expected, this is perhaps the first experiment to demonstrate interference for a genuine single photon state, as GRA themselves have emphasized.

\section{GRA's experiments according to CIEM}
GRA concluded from their results that  in a which-path measurement  a photon does not split at the beam-splitter and therefore chooses only one path, but, in a one-photon-at-a-time  interference experiment a photon splits at the beam-splitter and interferes with itself to produce an interference pattern. They view this result as experimental confirmation of particle-wave duality, and hence, of Bohr's principle of complementarity.

Without doubt, GRA's experiments with the novel and ingenious gating system constitutes an important experimental confirmation of quantum mechanics for genuine single photon states. But, by providing a detailed wave model of both experiments, we want to show that GRA's experiments cannot be regarded as confirmation of particle-wave duality, and hence, nor of Bohr's principle of complementarity.

We refer the reader to reference \cite{K87}, but particularly reference \cite{K94} for details of CIEM. Before proceeding we first give an outline of CIEM as given in reference (\cite{K05}, p. 300). 

\subsection{Outline of CIEM}
In what follows we use the radiation gauge in which the divergence of the vector potential is zero $\nabla.\Ab(\xxb,t)=0$, and the scalar potential is also zero $\phi(\xxb,t) = 0$. In this gauge the electromagnetic field has only two transverse components. Heavyside-Lorentz units are used throughout.

Second quantization is effected by treating the field $\Ab(\xxb,t)$ and its conjugate momentum $\Pb(\xxb,t)$ as operators satisfying  the equal-time commutation relations. This procedure is equivalent to introducing a field Schr\"{o}dinger equation
\begin{equation}
\int {\cal H} ( \Ab', \Pb') \fA\; d\xxb'= i \hbar \frac{\partial \fA}{\partial t},\label{SE}
\end{equation}
where the Hamiltonian density operator ${\cal H}$ is obtained from the classical Hamiltonian density of the  electromagnetic field,
\begin{equation}
{\cal H} =\frac{1}{2}(\mbox{\boldmath{$E$}}^{2}+\mbox{\boldmath{$B$}}^{2})=
\frac{1}{2}[c^{2}\Pb^{2}+(\nabla\times\Ab)^2 ], \label{H}
\end{equation}
by the  operator replacement $\Pb\rightarrow -i\hbar\, \delta'/\delta' \Ab$. $\Ab'$ is shorthand for $\Ab(\xxb',t)$. In earlier articles \cite{K05, K94} $\delta'/\delta' \Ab$ (without the prime) was defined as the variational derivative \footnote{\label{FD} For a scalar function $\phi$ the variational or functional derivative is defined  as $\frac{\delta'}{\delta' \phi}=\frac{\partial}{\partial\phi}-\Sigma_i\left(\frac{\partial}{\partial\left(\frac{\partial \phi}{\partial x_i}\right)}\right)$ (\cite{SHFF68},  p. 494). For a vector function $\Ab$ we have defined it to be $\frac{\delta}{\delta \Abs}=\frac{\delta}{\delta A_{x}}{\mbox{{\boldmath{$\mit i$}}}}+\frac{\delta}{\delta A_{y}}{\mbox{{\boldmath{$\mit j$}}}}+\frac{\delta}{\delta A_{z}}{\mbox{{\boldmath{$\mit k$}}}}$, where each component is defined in the same as for the scalar function.}. This definition leads to the equal-time commutation relations 
\[
[A_{i}(\xxb,t), \mathit{\Pi}_{j}(\xxb',t)]=-\frac{1}{c}[A_{i}(\xxb,t), E_{j}(\xxb',t)]=  i\hbar \delta_{ij}\delta^3( \xxb - {\xxb'}).
\]
Unfortunately, these commutation relations are known to be inconsistent both with  Gauss's law in free space, $\nabla.\mbox{\boldmath{$E$}}=0$, and the Coulomb gauge condition, $\nabla.\Ab=0$, since it follows from these that either of the two left-hand-side terms are zero, whereas the divergence of the delta function $\delta^3( \xxb - {\xxb'})$ is not zero. We noted this inconsistency in our original development of CIEM \cite{K94}, but justified this simplification by noting that it leads to the correct equations of motion. This justification, however, has recently been criticized by Struyve in reference \cite{WS2005}, p. 88\footnote{We would like to thank one of the referees for pointing out this reference and for re-emphasizing this inconsistency.}. As is well known, the commutation relations that are consistent with $\nabla.\mbox{\boldmath{$E$}}=0$  and $\nabla.\Ab=0$ are
\begin{equation}
[A_{i}(\xxb,t), \mathit{\Pi}_{j}(\xxb',t)]=i\hbar \delta_{ij}^{tr}( \xxb - \xxb') \label{CCB}
\end{equation}
where $ \delta_{ij}^{tr}( \xxb - \xxb')$ is the transverse delta function defined by \cite{BJD, LHR}
\[
 \delta_{ij}^{tr}(\xxb-\xxb')= \frac{1}{(2 \pi)^3} \int e^{i{\mbox{{\boldmath{$\mit k$}}}}.(\xxb-\xxb')}
\left ( \delta_{ij} - \frac{k_i k_j}{\mbox{{\boldmath{$\mit k$}}}^{2}} \right ) d^3{\mbox{{\boldmath{$\mit k$}}}}=\left(\delta_{ij} - \frac{\partial_i \partial_j}{\nabla^2}\right)\delta^3(\xxb-\xxb')
\]
We can establish consistency with the correct equal-time commutation relations, eq. (\ref{CCB}), by modifying the definition of the momentum operator as follows:
\[
\mathit{\Pi}_i=-i\hbar\frac{\delta}{\delta A_i}=-i\hbar\left( \frac{\delta'}{\delta' A_i}-\sum_k\frac{\partial_i\partial_k}{\nabla^2}\frac{\delta'}{\delta' A_k}\right),
\]
where $\delta'/\delta' A_k$ is the usual functional derivative defined in footnote \ref{FD}.

We note that the definition of the normal mode momentum operator given in the original article in which CIEM is developed \cite{K94} is consistent with the correct commutation relations, eq. (\ref{CCB}), and does not need modification.  

The solution of the field Schr\"{o}dinger  equation is the wave functional $\fA$. The square of the modulus of the wave functional $|\fA|^2$ gives the probability density  for a given field configuration $\Ab(\xxb,t)$. This suggests that we take $\Ab(\xxb,t)$ as a beable. Thus, as  we have already said, the basic ontology is that of a field; there are no photon particles.

We substitute $\Fi=R[\Ab,t]\exp(iS[\Ab,t]/\hbar)$, where $R[\Ab,t]$ and $S[\Ab,t]$ are two real functionals which codetermine one another, into the field Schr\"{o}dinger equation. Then,  differentiating, rearranging and equating imaginary  terms gives a continuity equation:

\begin{equation}
\frac{\partial R^{2}}{\partial t} + c^{2} \int \frac{\delta}{\delta \Ab'}
\left(R^{2}\frac{\delta S}{\delta \Ab'} \right) \; d\xxb' = 0.
\end{equation}
The  continuity  equation is interpreted as expressing conservation of probability in function space. Equating real terms gives a Hamilton-Jacobi type equation:
\begin{equation}
\frac{\partial S}{\partial t}+\frac{1}{2}\int\left(\frac{\delta S}{\delta
\Ab'}\right)^{2} c^{2}+(\nabla\times\Ab')^{2}+\left(-\frac{\hbar^2
c^{2}}{R}\frac{\delta^{2} R}{\delta \Ab'^{2}} \right) d\xxb'= 0. \label{HJ1}
\end{equation}
This Hamilton-Jacobi equation  differs from its classical counterpart by the extra classical term 
\begin{equation}
Q =-\frac{1}{2}\int\frac{\hbar^{2} c^{2}}{R}
\frac{\delta^{2} R}{\delta \Ab'^{2}}\;d\xxb',
\end{equation}   
which we call the field quantum potential.

By analogy with classical Hamilton-Jacobi theory we define  the total energy and  momentum
conjugate to the field  as
\begin{equation}
E = -\frac{\partial S[\Ab]}{\partial t},\;\;\;\;\;\Pb=\frac{\delta S[\Ab]}{\delta \Ab}.
\end{equation}
In addition to the beables $\Ab(\xxb,t)$ and $\Pb(\xxb,t)$, we can define  other field beables: the electric field, the magnetic induction, the energy and energy density, the momentum and momentum density, the intensity, etc. Formulae for these beables are obtained by replacing $\Pb$ by $\delta S/\delta \Ab$ in the classical formula. 

Thus,  we can picture an electromagnetic field as a field in the classical sense, but with the additional
property of  nonlocality. That the field is inherently nonlocal, meaning that an interaction at
one point in the field instantaneously influences the field at all other points, can be seen in
two ways: First, by using Euler's method of finite differences a functional can be
approximated as a  function of infinitely many variables:
$\fA\rightarrow\Fi(\Ab_1,\Ab_2,\ldots,t)$.  Comparison with a many-body wavefunction
$\psi(\xxb_1,\xxb_2,...,t)$ reveals the nonlocality.   The second way  is from the equation of motion
of $\Ab(\xxb,t)$, i.e.,  the free field wave equation. This is obtained by taking the  functional
derivative of the Hamilton-Jacobi equation, (\ref{HJ1}):
\begin{equation}
\nabla^{2}\Ab-\frac{1}{c^{2}}\frac{\partial^{2}\Ab}{\partial t^{2}}= \frac{\delta
Q}{\delta\Ab}.
\end{equation}
In general $\delta Q/\delta\Ab$ will involve an integral over space in which the
integrand contains $\Ab(\xxb,t)$. This means that the way that $\Ab(\xxb,t)$ changes with time at
one point depends on $\Ab(\xxb,t)$ at all other points, hence the inherent nonlocality.

\subsection{Normal mode coordinates\label{NMC}}

To proceed it is mathematically easier to expand $\Ab(\xxb,t)$ and $\Pb(\xxb,t)$ as Fourier series
\begin{equation}
\Ab(\xxb,t)=\frac{1}{\V}\sk\ek\qk(t)\ep,\;\;\;\;\;\;\;\;\;\;
\Pb(\xxb,t)=\frac{1}{\V}\sk\ek\pk(t) \epm, \label{AFS}
\end{equation}
where the field is assumed to be enclosed in a large volume $V=L^3$. The wavenumber $k$ runs from $-\infty$ to $+\infty$ and $\mu=1,2$ is the polarization index. For $\Ab(\xxb,t)$ to be a real function we must have 
\begin{equation}
\mbox{\boldmath $\hat\varepsilon$}_{-k\mu}q_{-k\mu}=\ek\qks.\label{QMP}
\end{equation}
Substituting eq.'s (\ref{H}) and (\ref{AFS})  into eq. (\ref{SE})  gives the Schr\"{o}dingier equation in terms of the normal mode coordinates $\qk$:
\begin{equation}
\frac{1}{2}\sk\left(-\hbar^{2}c^{2}\frac{\partial^2\Fi}{\partial
\qks\partial\qk }+\kappa^{2}\qks\qk\Fi\right)=
i\hbar\frac{\partial\Fi}{\partial t}. \label{SEN}
\end{equation}
The solution   $\Fi( \qk,t)$ is an ordinary
function of all the normal mode coordinates and this simplifies proceedings. 

We substitute  $\Fi=R(\qk,t)\exp[iS(\qk,t)/\hbar]$, where $R(\qk,t)$  and $S(\qk,t)$ are real functions which codetermine one another, into eq. (\ref{SEN}). Then,  differentiating, rearranging and equating real terms  gives the continuity  equation in terms of normal modes:
\begin{equation}
\frac{\partial R^2}{\partial t}+\sk\left[\frac{c^2}{2}\frac{\partial}{\partial
\qk}\left(R^2\frac{\partial S}{\partial\qks}\right)+
\frac{c^2}{2}\frac{\partial}{\partial\qks}\left(R^2\frac{\partial S}{\partial\qk}
\right) \right]=0.
\end{equation}
 Equating imaginary terms gives the Hamilton-Jacobi equation in terms of normal modes:
\begin{equation}
\frac{\partial S}{\partial t}+\sk\left[\frac{c^2}{2}\frac{\partial
S}{\partial\qks}\frac{\partial S}{\partial \qk}+\frac{\kappa^{2}}{2}\qks\qk
+\left(-\frac{\hbar^{2} c^{2}}{2R}\frac{\partial^{2}R}
{\partial\qks\partial\qk}\right)\right]=0. \label{HJ2}
\end{equation}
The term 
\begin{equation}
Q = -\sk\frac{\hbar^{2} c^{2}}{2R}\frac{\partial^{2}R} {\partial\qks\partial\qk} \label{QP}
\end{equation}
is the field quantum potential. Again, by analogy with classical Hamilton-Jacobi theory we define the total energy and the conjugate momenta as
\begin{equation}
E=-\frac{\partial S}{\partial t},\;\;\;\;\;\pk=\frac{\partial S}{\partial\qk},\;\;\;\;\;
\pks=\frac{\partial
S}{\partial\qks}.
\end{equation}
The square of the modulus of the wave function $|\Fi( \qk,t)|^2$  is  the probability density
for each $\qk(t)$ to take a particular value at time $t$.  Substituting a particular set of values of $\qk(t)$ at time $t$ into eq. (\ref{AFS}) gives a particular field  configuration at time $t$, as before. Substituting  the initial values of $\qk(t)$ gives the initial field configuration. 

The normalized ground state solution of the Schr\"{o}dinger equation is given by
\begin{equation}
\Fi_0=N e^{-\sum_{k\mu}(\kappa/2\hbar c)q_{k\mu}^{*}q_{k\mu}}e^{-\sum_{k}i\kappa ct/2},
\end{equation}
with $N= \prod_{k=1}^{\infty}(k/\hbar c \pi)^{\frac{1}{2}}$\footnote{The normalization factor $N$ is found by substituting $\qks=f_{k\mu}+ig_{k\mu}$ and its conjugate into $\Fi_0$ and using the normalization condition $\int_{-\infty}^{\infty} |\Fi_0|^2df_{k\mu}dg_{k\mu}=1$, with $df_{k\mu}\equiv df_{k_11}df_{k_12}df_{k_21}\ldots$, and similarly for $dg_{k\mu}$.}. Higher excited states are obtained by the action of the creation operator $a^{\dag}_{k\mu}$:
\begin{equation}
\Fi_{n_{k\mu}}=\frac{(a_{k\mu}^{\dag})^{n_{k\mu}}}{\sqrt{n_{k\mu}!}}\Fi_{0}e^{- in_{k\mu}\kappa ct}.
\end{equation}
For a normalized ground state, the higher excited states remain normalized. For ease of writing we will not include the normalization factor $N$ in most expressions, but normalization of states will be assumed when calculating expectation values.

Again, the formula for the field beables are obtained by replacing the conjugate momenta $\pk$ and $\pks$ by $\partial S/\partial\qk$ and $\partial S/\partial\qks$ in the corresponding classical formula. The following is a list  of formulae for the beables:\\ \mbox{}\\

The vector potential $\Ab(\xxb,t)$ is given in eq. (\ref{AFS}). The electric field is
\begin{equation}
\E(\xxb,t)=-c\Pb(\xxb,t)=-\frac{1}{c}\frac{\partial \Ab}{\partial t}= - \frac{c}{\V}\sk\ek\frac{\partial S}{\partial\qk}\epm. \label{PEX}
\end{equation}
The magnetic induction is
\begin{equation}
\B(\xxb,t)=\nabla\times\Ab(\xxb,t)=\frac{i}{\V}\sk(\kk\times\ek)\qk(t)\ep. \label{BEX}
\end{equation}
We may also define the energy density, which includes the quantum potential density (see reference \cite{K94}), but we will not write these here as we will not need them.  The total energy is found by integrating the energy density over $V$ to get
\begin{equation}
E=-\frac{\partial S}{\partial t}=\sk\left[\frac{c^{2}}{2} \frac{\partial
S}{\partial\qks}\frac{\partial S}{\partial\qk}+\frac{\kappa^{2}}{2}\qks\qk
+\left(-\frac{\hbar^{2}c^{2}}{2R}\frac{\partial^{2}R} {\partial
\qks\partial\qk} \right)\right]. 
\end{equation}
The intensity is equal to momentum density multiplied by $c^2$:
\begin{equation}
\I(\xxb,t)=c^2\mbox{\boldmath${\cal G}$}= \frac{-ic^2}{V}\sk\sum_{k'\mu'}\left[ \ekp\times(\kk\times\ek)\frac{\partial S}{\partial q_{k'\mu'}}\qk\epp \right]. \label{I}
\end{equation}

We have adopted the classical definition of intensity in which the intensity is equal to the Poynting vector (in Heavyside-Lorentz units), i.e., $\I=c(\E\times\B)$. The definition leads to  a moderately simple formula for the intensity beable.  We note that the definition above contains a zero point intensity. But, because $\I$ is a vector (whereas energy is not) the contributions to the zero point intensity from individual waves with wave vector $\kk$ cancel each other because of symmetry; for each $\kk$ there is another $\kk$ pointing in the opposite direction. The above, however,  is not the definition normally used in quantum optics. This is probably because, although it leads to a simple formula for the intensity beable, it leads to a very cumbersome expression for the intensity operator in terms of the creation and annihilation operators:
\begin{eqnarray}
&&\mbox{{ \boldmath{$\Ih$}}}
=\frac{-\hbar c^2}{4V}\sk\sum_{k'\mu'}\left[ \frac{k}{k'}\ek\times(\kk'\times\ekp)- \frac{k'}{k}(\kk\times\ek)\times\ekp \right] \nonumber\\
&&\times\left[ \hat{a}_{k\mu}\hat{a}_{k'\mu'}e^{i(\kb+\kbp).\xb} -  \hat{a}_{k\mu}\hat{a}^\dag_{k'\mu'}\epp- \hat{a}^\dag_{k\mu}\hat{a}_{k'\mu'}\eppm +\hat{a}^\dag_{k\mu}\hat{a}^\dag_{k'\mu'}e^{-i(\kb+\kbp).\xb} \right].  \label{Ipan}
\end{eqnarray}
In quantum optics the intensity operator is defined instead as $\Ih
=c(\mbox{{ \boldmath{$\hat{E}^+$}}}\times \mbox{{ \boldmath{$\hat{B}^-$}}} - \mbox{{ \boldmath{$\hat{B}^-$}}}\times \mbox{{ \boldmath{$\hat{E}^+$}}})$, and leads to a much simpler expression in terms of creation and annihilation operators
\begin{equation}
\Ih= \frac{\hbar c^2}{V}\sk\sum_{k'\mu'}\hat{\kk}\sqrt{kk'} \hat{a}^\dag_{k\mu}\hat{a}_{k'\mu'}e^{i(\kbp-\kb).\xb}. \label{Iqo}
\end{equation}
This definition is justified because it is proportional to the dominant term in the interaction Hamiltonian for the photoelectric effect upon which instruments that measure intensity are based. We note that the two forms of the intensity operator lead to identical expectation values and perhaps further justifies the simpler definition of the intensity operator. 

From the above we see that objects such as $\qk$, $\pk$, etc., regarded as time independent operators in the Schr\"{o}dinger picture of the usual interpretation, become functions of time in CIEM. 

For a given state $\Fi(\qk,t)$ of the field we determine the beables by first finding $\partial S/\partial\qk$ and its complex conjugate using the formula 
\begin{equation}
S=\left(\frac{\hbar}{2i}\right)\ln\left(\frac{\Fi}{\Fi^*}\right).\label{FlaS}
\end{equation}
This gives the beables as functions  of the $\qk(t)$ and $\qks(t)$. The beables can then be obtained in terms of the initial values by solving the equations of motion for $\qk(t)$ and $\qks(t)$. There are two alternative but equivalent forms of the equations of motion. The first follows from the classical formula 
\begin{equation}
\pk =  \frac{\partial{\cal L}}{\partial\left(\frac{d \qk}{d t} \right)} =\frac{1}{c^2}\frac{d\qks}{d t}, 
\end{equation}
where ${\cal L}$ is the Lagrangian density of the electromagnetic field,  by replacing $\pk$ by $\partial S/\partial\qk$. This gives the equations of motion as
\begin{equation}
\frac{1}{c^2}\frac{d\qks(t)}{d t} =\frac{\partial S}{\partial\qk(t)}.\label{EQMG}
\end{equation}
The second form of the equations of motion for $\qk$ is obtained by differentiating the Hamilton Jacobi equation (\ref{HJ2}) by $\qks$. This gives the wave equations
\begin{equation}
\frac{1}{c^2}\frac{d^{2}\qks}{d t^{2}}+\kappa^{2}\qks=-\frac{\partial
Q}{\partial\qk}. \label{WEQ}
\end{equation}
The corresponding equations   for $\qk$ are the complex conjugates of the above. These equations of motion  differ from the classical free field wave equation by the derivative of the quantum potential. From this it follows that where the quantum potential is zero or small the quantum field  behaves like a classical field. In applications we will obviously choose to solve the simpler eq. ({\ref{EQMG}}). 

We conclude with a few words to clarify our model. The electromagnetic field beables are  $\E(\xxb,t)$ and $\B(\xxb,t)$ and are objectively existing entities in real space. The state $\Fi=R\exp[iS/\hbar]$ is made up of the $R$ and $S$ functionals. By thinking in terms of the approximation of a functional as a function of infinitely many variables or in term of normal mode coordinates we can picture  $R$ and $S$ as connecting the field coordinates and shaping the behaviour of the field through the equations of motion 
(\ref{EQMG}) or (\ref{WEQ}), but the $R$ and  $S$ beables (and hence the state $\Fi$) are not the electromagnetic field itself.  The $R$ and $S$ beables co-determine one another and the motion of the field can be determined from either one without reference to the other. This is reflected in the two possible forms of the equations of motion. 

\subsection{GRA's which-path experiment according to CIEM}
Refer to figure \ref{GRA1}.  To keep the mathematics simple we assume (a) a symmetrical beam-splitter so that the reflection and transmission coefficients are equal and given by $r=t=1/\sqrt{2}$, (b) a $\pi/2$ phase shift upon reflection, and (c) no phase shift upon transmission. With this in mind, the state of the photon after the beam-splitter but before the mirrors and phase shifter is
\begin{equation}
\Fi_{I}=\frac{1}{\sqrt{2}}\left( \Fi_{\alpha}+i\Fi_{\beta}\right), \label{PHRI}
\end{equation}
where $\Fi_{\alpha}$ and $\Fi_{\beta}$ are solutions of the normal mode Schr\"{o}dinger equation and are given by
\begin{eqnarray}
\Fi_{\alpha}(\qk,t)& =& \left(\frac{2\kappa_\alpha}{\hbar c}\right)^{\frac{1}{2}}
\aks\Fi_{0}e^{-i\kappa_\alpha ct}, \;\;\;\;\;\;\;\;\;
\Fi_{\beta}(\qk,t) = \left(\frac{2\kappa_\beta}{\hbar c}\right)^{\frac{1}{2}}
\bks\Fi_{0}e^{-i\kappa_\beta ct},\nonumber\\
\Fi_0(\qk,t)&=&N e^{-\sum_{k\mu}(\kappa/2\hbar c)q_{k\mu}^{*}q_{k\mu}}e^{-\sum_{k}i\kappa ct/2}. 
\end{eqnarray}
The magnitudes of the $k$-vectors are equal, i.e., $k_\alpha=k_\beta=k_0$. The $\ak$ normal mode coordinates represent the horizontal beam and the $\bk$ coordinates represent the vertical beam. It is clear that the single photon input state $\Fi_i(\qk,t) =(2\kappa_0/(\hbar c))^{\frac{1}{2}}
q_{k_0\mu_0}^{*}(t)\Fi_{0}e^{-i\kappa_0 ct}  $ is split by the beam-splitter into two beams. This remains true irrespective of whether a subsequent measurement is a which-path measurement or it is the observation of interference. The mathematical description is unique.

In CIEM the normal mode coordinates are regarded as functions of time and represent an actually existing  electromagnetic field. The modulus squared of the wavefunction is a probability density from which the probabilities for the normal modes to have particular values are found. The totality of these probabilities gives the probability for a particular field configuration. Thus, the ontology is that of a field; there are no photon particles. In fact, for a number state  the most probable field configuration is one or more plane waves, which, in general, are nonlocal (\cite{K94}, p. 326). As we mentioned earlier, in CIEM we use the term photon to refer to  a quantum of energy $\hbar\omega$ (or an average about this value for a wave packet) without in any way implying particle properties.

To find the equations of motion for the normal mode coordinates we first find $S$ from $\Fi_{I}=R(\qk,t)\exp(iS(\qk, t))$ and then substitute into 
\begin{equation}
\frac{1}{c^2}\frac{d\qks(t)}{d t} =\frac{\partial S}{\partial\qk(t)}.
\end{equation}
This gives the equations of motion
\begin{eqnarray}
\frac{d\aks}{dt}&=&c^2\frac{\partial S}{\partial\ak}=\frac{\hbar c^2}{2}\frac{i}{\left(\ak-i\bk \right)}, \label{EQMa} \\
\frac{d\bks}{dt} &=&c^2\frac{\partial S}{\partial\bk} =\frac{\hbar c^2}{2}\frac{1}{\left(\ak-i\bk \right)}, \label{EQMb}\\
\frac{d\qks}{dt} &=&c^2\frac{\partial S}{\partial\qk}  =0,\;\;\;\; \mathrm{for}\; k\neq \pm k_{\alpha}, \pm k_{\beta}.  \label{EQMq}
\end{eqnarray}
Eqs. (\ref{EQMa}) and (\ref{EQMb}) are coupled differential equations and the coupling indicates that the two beams are nonlocally connected. The solutions are
\begin{equation}
\aks(t)=\alpha_0 e^{i(\omega_\alpha t+\sigma_0)}, \;\;\;
\bks(t)=\beta_0 e^{i(\omega_\beta t+\tau_0)}, \;\;\;
\qks(t)= q_{k\mu 0}e^{i\zeta_{k\mu 0}}\;\mathrm{for}\;k\neq \pm k_{\alpha}, \pm k_{\beta}, \label{QKS}
\end{equation}
where $\sigma_0$ and $\tau_0$ are integration constants corresponding to the initial phases, and $\alpha_0$ and $\beta_0$ are constant initial amplitudes. The omega's, $\omega_\alpha=\hbar  c^2/4\alpha_0^2$ and $\omega_{\beta}=\hbar c^2/4\beta_0^2$, are nonclassical frequencies which depend on the amplitudes $\alpha_0$ and $\beta_0$. The vector potential, electric intensity, magnetic induction and intensity beables are given by the formulae
\begin{eqnarray}
\Ab(\xxb,t)&=&\frac{1}{\V}\sk\ek\qk(t)\ep,\nonumber\\
\E(\xxb,t)&=&-c\Pb(\xxb,t)=-\frac{1}{c}\frac{\partial \Ab}{\partial t}= - \frac{c}{\V}\sk\ek\frac{\partial S}{\partial\qk}\epm, \nonumber\\
\B(\xxb,t)&=&\nabla\times\Ab(\xxb,t)=\frac{i}{\V}\sk(\kk\times\ek)\qk(t)\ep, \nonumber\\
\I(\xxb,t)&=&c^2\mbox{\boldmath${\cal G}$}= \frac{-ic^2}{V}\sk\sum_{k'\mu'}\left[ \ekp\times(\kk\times\ek)\frac{\partial S}{\partial q_{k'\mu'}}\qk\epp \right].
\end{eqnarray}
Substituting equations (\ref{EQMa}) to (\ref{QKS}) into the above formulae gives the field beables  associated with the state $\Fi_I$:
\begin{eqnarray}
\Ab_I(x,t)&=&\frac{2}{\V}\left(\eka\alpha_0\cos\Th_{\alpha}+\ekb\beta_0\cos\Th_{\beta}
\right) +\frac{\ux_I(\xxb)}{\V},\nonumber \\ 
\E_I(\xxb,t)&=&\frac{-\hbar c}{2\V}\left(\frac{\eka}{\alpha_0}\sin\Th_{\alpha}+\frac{\ekb}{\beta_0}\sin\Th_{\beta} \right), \label{EI}\\
\B_I(\xxb,t) &=&\frac{-2}{\V}\left[(\ka\times\eka)\alpha_0\sin\Th_{\alpha} + (\kkb\times\ekb)\beta_0 \sin \Th_{\beta}\right]  + \frac{\vx_I(\xxb)}{\V}, \nonumber\\ 
\I_I(\xxb,t)&=&\frac{\hbar c^2}{2V}\left(\ka+\kkb-\ka\cos2\Th_{\alpha} -\kkb\cos 2\Th_{\beta}\right)-\frac{\ffx_I(\xxb)\gx_I(\xxb,t)}{V}, 
\end{eqnarray}
with $\Th_{\alpha}=\ka.\xxb-\omega_{\alpha} t-\sigma_0$ and $\Th_{\beta}=\kkb.\xxb-\omega_{\beta} t-\tau_0$, and 
\begin{eqnarray}
\ux_I(\xxb)&=&\!\!\!\!\!\sum_{\stackrel{\scriptstyle{k\mu}}{k\neq \pm k_{\alpha},\pm k_{\beta}}}\!\!\!\!\!\ek\qk\ep, 
\;\;\;\;\;\vx_I(\xxb)=\nabla\times\ux_I(\xxb)=i\!\!\!\!\!\sum_{\stackrel{\scriptstyle{k\mu}}{k\neq \pm k_{\alpha},\pm k_{\beta}}}\!\!\!\!\!(\kk\times\ek)\qk\ep, \label{VXI}\\
\ffx_I(\xxb)&=&i\hbar c^2\!\!\!\!\!\sum_{\stackrel{\scriptstyle{k\mu}}{k\neq \pm k_{\alpha},\pm k_{\beta}}}\!\!\!\!\!\eka\times(\kk\times\ek)\qk\ep,\;\;\;\;\;
\gx_I(\xxb,t)=\sin \Th_{\alpha}+\sin \Th_{\beta}.
\end{eqnarray}

Complementarity is not a direct  interpretation of the mathematical formalism, so that the uniqueness of the mathematical description is not reflected in the duality of complementary concepts. The ontology of CIEM, on the other hand, is a direct interpretation of the elements of the mathematical formalism. The beables above therefore reflect the splitting of the state $\Fi_i$ into two beams. In other words, the photon always splits at the beam-splitter irrespective of the nature of any planned future measurement.

Quantum mechanics predicts that in a which-path measurement  a photon will be detected in only one path. Feeble light experiments of the past have confirmed this prediction indirectly, while GRA's which-path experiment  provides direct confirmation.  Our CIEM model must therefore explain how a photon is detected in only one path, even though the photon must split at the beam-splitter. To see how this comes about we outline the interaction of the electromagnetic field in state $\Fi_I$ with the photomultipliers. For mathematical simplicity we model the photomultipliers $PM_t$ and $PM_r$ as hydrogen atoms. We assume that the incident photon has sufficient energy to ionize one of the  hydrogen atoms.

The treatment we give here is a short summary of a more detailed outline given in reference (\cite{K05}, p. 310). The initial state of the field before interaction with the hydrogen atom is given by eq. (\ref{PHRI}). The initial state of the hydrogen atom is 
\begin{equation}
u_i(\xxb,t)=\frac{1}{\sqrt{\pi a^3}}e^{-r/a}e^{-iE_{ei}t/\hbar},
\end{equation}
where $a=4\pi\hbar^2/\mu e^2$ is the Bohr magneton.  With the initial state $\Fi_{I_{k\mu}i}(\qk,\xxb,t)=\Fi_{I_{k\mu}}(\qk,t)u_i(\xxb,t)$, the Schr\"{o}dinger equation
\begin{equation}
i\hbar\frac{\partial \Fi}{\partial t}= (H_R+H_A+H_I)\Fi
\end{equation}
can be solved using standard perturbation theory. $H_R$, $H_A$ and $H_I$ are the free radiation, free atomic, and interaction Hamiltonians, respectively, and are given by
\begin{equation}
H_R = \sk\left(a^{\dag}_{k\mu}a_{k\mu}+\frac{1}{2}\right)\hbar\omega_k, \;\;\;\;\; H_A = \frac{-\hbar^2}{2\mu}\nabla^2+V(\xxb), \;\;\;\;\;H_I = \frac{i\hbar e}{\mu c}\left(\frac{\hbar c}{2V}\right)^{\frac{1}{2}} \sk\frac{1}{\sqrt{k}}a_{k\mu}\ep\ek.\nabla,
\end{equation}
with $\omega_k=kc$ and $\mu=m_em_n/(m_e+m_n)$ is the reduced mass. The final solution is 
\begin{equation}
\Fi=\Fi_{I_{k\mu}i}(\qk,\xxb,t)-\frac{\Fi_0(\qk,t)}{V}\sum_n \eta_{0n}(t)\eko.\kk_{en}\frac{1}{\sqrt{V}}e^{i\left(\kb_{en}.\xb-E_{en}t/\hbar\right)}, \label{FSOL}
\end{equation}
with
\begin{equation}
\eta_{0n}(t)=\left(\frac{e}{\mu c}\right)\sqrt{\frac{\hbar c}{2V}}\left[\frac{(i-e^{i\phi})}{\sqrt{2 k_0}}\right]\left[\frac{\hbar}{\sqrt{V\pi a^3}} \frac{8\pi a^3}{(1+a^2 k_{en}^2)^2} \right]\left(\frac{1- e^{iE_{0n,I_{k\mu}i}t/\hbar}}{E_{0n,I_{k\mu}i}}\right).
\end{equation}
$E_{0n,I_{k\mu}i}$ is given by 
\begin{equation}
E_{0n,I_{k\mu}i}=E_0+E_{en}-E_{I_{k\mu}}-E_{ei}.
\end{equation}
Eq. (\ref{FSOL}) clearly shows that one entire photon is absorbed. This is further emphasized by the integral
\begin{equation}
\sum_{k\mu}\frac{1}{\sqrt{k}}\int\Fi^{*}_{N_{k\mu}}a_{k\mu}\Fi_{I_{k\mu}}\;d\qk 
= \frac{1}{\sqrt{2k_0}}(i-e^{i\phi})\int \Fi^{*}_{N_{k\mu}}\Fi_0\;d\qk= \frac{1}{\sqrt{2k_0}}(i-e^{i\phi})\delta_{N_{k\mu}0}\delta_{kk_0}\delta{\mu\mu_0},
\end{equation}
which is part of the matrix element $H_{N_{k\mu}n,I_{k\mu}i}$ used in obtaining the final solution. This term shows that if the interaction takes place at all then an entire  electromagnetic quantum must be absorbed by the hydrogen atom.

The initial state $\Fi_{I_{k\mu}}$ represents a single photon divided between the two beams, but in the interaction with an atom positioned in one of the beams, the entire photon must be absorbed. Given that the interferometer arms can be of arbitrary  length such absorption must in general be nonlocal. In this way we can explain why a photon that always divides at the beam-splitter nevertheless registers in only one path. The fact that this wave model exists prevents GRA's which-path experiment from being regarded as confirmation of the particle behaviour of light.  

\subsection{GRA's interference experiment according to CIEM}
Refer to figure \ref{GRAmz}. Using the same phase and amplitude changes as in the previous section, and tracing the development of the two beams after $BM_2$, we arrive at the wavefunction
\begin{equation}
\Fi_{II}=-\frac{1}{2}\Fi_{c}(1+e^{i\phi})+\frac{i}{2}\Fi_d(1-e^{i\phi}).
\end{equation}  
By following a similar procedure to that of region I, we can find the $S$ corresponding to $\Fi_{II}$ and hence set up and solve the equations of motion. Using these solutions the beables for region II are found to be
\begin{eqnarray}
\Ab_{II}(\xxb,t)&=&\frac{2}{\V}\left( \ekc c_0 \cos\Th_c +\ekd d_0\cos \Th_d\right)+\frac{\ux_{II}(\xxb)}{\V},\nonumber\\
\E_{II}(\xxb,t)&=&\frac{-\hbar c}{2\V}\left(\frac{\eka}{c_0}(1+\cos\phi)\sin\Th_c + \frac{\ekd}{d_0}(1-\cos\phi)\sin\Th_d \right),\nonumber\\
\B_{II}(\xxb,t) &=&\frac{-2}{\V}\left[(\kc\times\ekc)c_0\sin\Th_c + (\kd\times\ekd)d_0 \sin\Th_d\right] +\frac{\vx_{II}(\xxb)}{\V}, \nonumber\\
\I_{II}(\xxb,t)&=&\frac{\hbar c^2}{2V}\left[\kc(1+\cos\phi)+\kd(1-\cos\phi) -\kc(1+\cos\phi)\cos2\Th_c+ \kd(1-\cos\phi)\cos 2\Th_d)\right] \nonumber\\
&&-\frac{\ffx_{II}(\xxb)\gx_{II}(\xxb,t)}{V}, \label{INTII}
\end{eqnarray}
with
\begin{eqnarray}
\ux_{II}(\xxb)&=&\!\!\!\!\!\sum_{\stackrel{\scriptstyle{k\mu}}{k\neq \pm k_c,\pm k_d}}\!\!\!\!\!\ek\qk\ep,\;\;\;\;\;
\vx_{II}(\xxb)=i\!\!\!\!\!\sum_{\stackrel{\scriptstyle{k\mu}}{k\neq \pm k_c,\pm k_d}}\!\!\!\!\!(\kk\times\ek)\qk\ep=\nabla\times \ux_{II}(\xxb), \label{VXII}\\
\ffx_{II}(\xxb)&=&\frac{i\hbar c^2}{V}\!\!\!\!\!\sum_{\stackrel{\scriptstyle{k\mu}}{k\neq \pm k_c,\pm k_d}}\!\!\!\!\!\eko\times(\kk\times\ek)\qk\ep,\; \gx_{II}(\xxb,t)=(1+\cos\phi)\sin\Th_c \\ \nonumber
&&\;\;\;\;\;\;\;\;\;\;\;\;\;\;\;\;\;\;\;\;\;\;\;\;\;\;\;\;\;\;\;\;\;\;\;\;\;\;\;\;\;\;\;\;\;\;\;\;\;\;\;\;\;\;\;\;\;\;\;\;\;\;\;\;\;\;\;\;\;\;\;\;\;\;\;\;\;\;\;\;\;\;+(1-\cos\phi)\sin\Th_d,
\end{eqnarray}
and with $\Th_c=\kc.\xxb-\omega_c t-\chi_0$ and $\Th_d=\kd.\xxb-\omega_d t -\xi_0$.

The wavefunction and the beables clearly show interference. For example, for $\phi=0$ the $d$-beam is extinguished and for $\phi=\pi$ the $c$-beam is extinguished by interference.

\section{Comments on some other recent experimental tests of complementarity\label{CGBAS}}

In the proposed experiment of Ghose {\it et al} \cite{GHOSE91}, light is incident on a prism at an angle greater than the critical angle and hence undergoes total internal reflection. A second prism placed less than a wavelength from the first  allows light to tunnel into the transmitted channel.  Quantum mechanics predicts perfect anticoincidence. This is interpreted by Ghose {\it et al}, as is usual, as which-path information and hence as particle behaviour. Transmitted photons necessarily tunnel through the gap between the prisms, a phenomenon which the authors interpret as wave behaviour. In this way, the authors claim that wave and particle behaviour are observed in the same experiment in contradiction to Bohr's principle of complementarity.  This experiment has since been performed by Mizobuchi {\it et al} \cite{MIZ92} using a GRA single photon source, but as we mentioned earlier, the statistical accuracy of their results has been questioned in references \cite{UNNIK, GHOSE99, BRIDA04}. 

To resolve the technical difficulties with Mizobuchi {\it et al}'s experiment, Brida {\it et al}, following a suggested experiment by Ghose \cite{GHOSE99} and also employing the GRA single photon source, used a birefringent crystal to split a light beam into two beams (the ordinary and the extraordinary beams) instead of  using tunneling between two closely spaced prisms.  They interpreted the birefringent splitting as wave behaviour, while the perfect anticoincidence they observed they interpreted as particle behaviour. Again, the claim is the observation of wave and particle behaviour in the same experiment in contradiction of complementarity.

Afshar's experiment is of the two-slit type. He first observes interference a short distance in front of the slits and determines the position of the dark fringes. He then replaces the screen with a wire grid such that the grid wires coincide with the dark fringes. A lens is placed after the grid to form an image of the two slits. The images showed no loss of sharpness or intensity as compared to the image of the two slits without the grid in position. Afshar concluded that there was interference prior to formation of the image which he interpretes as wave behaviour. He assumes that the images of the slits are formed by photons coming from the slit on the same side as the image. He then  interpretes image formation as providing path information, and hence particle behaviour. Ashar concludes that particle and wave  behaviour is observed in the same experiment in contradiction of complementarity. 

We do not agree that these experiments either disprove Bohr's principle of complementarity, or, as argued by Brida {\it et al},that they can be viewed as  a generalization of Bohr's principle of complementarity. Our reasons follow.

As for the GRA experiments, all the above experiments can be explained using CIEM, i.e., they can be explained entirely in terms of a wave model. One is therefore not forced  to conclude that these experiments require a generalization of Bohr's principle of complementarity (a generalization first suggested by Wootters and Zurek \cite{WZ} as mentioned in the introduction), a generalization  which is severely flawed, as mentioned in the introduction. We will comment further below. 

Arguments from the perspective of complementarity can be put  to show that these experiments do not disprove complementarity.  Let us first consider the experiments of Mizobuchi {\it et al} and  Brida {\it et al}. Bohr emphasized that only the final experimental result (pointer reading) has physical significance and that an experiment should be viewed as a whole, not further analyzable \cite{BR59A, BR28}. We recall the statement of Wheeler, `No phenomenon is a phenomenon until it is an observed phenomenon' (\cite{WHR78}, p 14). In these two experiments, the observed results are anticoincidence detections which the above authors and advocates of complementarity or its variants can reasonably and unambiguously attribute to particle behaviour. The wave behaviour is not detected. It is therefore perfectly consistent for a Bohrian to maintain that the experiments unambiguously define a particle model even if this is counter-intuitive. The Afshar experiment avoids this criticism because the presence of the wire grid physically detects the interference.  But, the Afshar experiment still fails because of the first point above, namely that CIEM  provides a wave model of image formation by a large series of single photon detections.  

Another point to consider is that the mutually exclusive wave and particle complementary concepts are not related to the mathematical formalism of the quantum theory. In this way they differ from complementary concepts such as position and momentum or the components of angular momentum which are not mutually exclusive classical concepts and are represented in the mathematical formalism of the quantum theory by Heisenberg uncertainty relations. In this case, what is called wave or particle behaviour in a given experiment is somewhat arbitrary. Apart from other points, this arbitrariness is an important reason why we feel complementarity can neither be proved nor disproved.

We now comment on a widely accepted  generalization  of complementarity  by Wootters and Zurek in their influential article \cite{WZ}. This generalization admits partial wave and partial  particle behaviour in the same experiment. Based on this generalization Wootters and Zurek \cite{WZ},  and later Yasin and Greenberger \cite{GY}, cast  particle-wave duality in mathematical form. We have argued in earlier articles \cite{KPW} that far from being a generalization of complementarity, this approach in fact contradicts complementarity.  From the mathematical perspective,  these mathematical relations are constructs appended to the  formalism of the quantum theory but not derived from it. As a measure of coherence they can be thought of as useful heuristic rules, but  for the reasons we will give,  can be attributed no more fundamental significance than this. For detailed arguments against this generalization we refer the reader to  reference \cite{KPW} and restrict ourselves here to briefly emphasizing aspects of complementarity which demonstrate our point of view. 

In his explanations of his principle of complementarity \cite{BR59A, JAM74,BR28}, Bohr repeatedly emphasized the mutual exclusiveness of complementary concepts, and the requirement of mutually exclusive experimental arrangements for their correct use or definition. He further emphasized that  complementary concepts are abstractions to aid thought, and cannot be attributed physical reality. It seems to the present author  that Bohr was concerned to provide a framework for the correct use of classical language or concepts.  Thus, for the same physical object to be both a wave and a particle is, quite simply, a contradiction of definitions. This, the present author believes, is what led Bohr to emphasize that complementary concepts could not be attributed physical reality. By insisting on mutually exclusive experimental arrangements for the realization of complementary concepts, Bohr, in the authors view, allowed for the use of classical language/concepts in a way that avoids contradiction. It is for these reasons that we regard the Wootters and Zurek generalization of complementarity in terms of partial particle behaviour/knowledge and partial wave beaviour/knowledge as the complete antithesis of Bohr's principle of complementarity. Even apart from Bohr's teachings, what can it mean for a physical object to be partially a wave and partially a particle? Above, we made a distinction between particle and wave complementary concepts and other pairs of complementary concepts that Bohr did not make. Our arguments here need not apply to complementary concepts such as position and momentum, which classically are not mutually exclusive concepts.  We note two things: First, the Wootters and Zurek generalization of complementarity is in terms of wave and particle concepts. Second, from the point of view of interpretation, particle and wave complementary concepts are the most fundamental, and lie at the heart of the interpretational issues of the quantum theory.

The experiment of Kim {\it et al} concerns both complementarity and the Wheeler delayed-choice issue, but its significance goes  beyond these issues. The results of this experiment appear to suggest that  a present measurement affects a past measurement. The Wheeler delayed-choice experiments indicate that a present measurement either creates or changes the past history leading to a particular result (there are subtle differences between Wheeler's and Bohr's position which are discussed in reference \cite{K05} section 1). The Kim  {\it et al}  and Wheeler delayed-choice experiment differ in that the past history is not actually observed in Wheeler's experiment, whereas in  Kim et al's experiment it is the result of an actual past measurement that is changed by a measurement in the present. We will leave a detailed discussion of this experiment for a later article, but make one observation. The experiment uses a pair of correlated photons produced by the process of spontaneous parametric down conversion. By detecting the photon partner {\it after} the first photon is detected, the earlier measured wave or particle behaviour of the first photon is determined. What seems to have been left out of the Kim {\it et al} analysis is that once the first photon is detected and the state of the EPR partner changes accordingly, thereafter, the EPR correlation is broken. Hence, any measurement  performed on the second photon can have no effect on its partner. This is a firm prediction of quantum mechanics. Nevertheless, the strange result in which a present measurement appears to determine the outcome of an earlier measurement needs explanation. Other articles relating to this issue can be found in reference \cite{QE}.
\section{Conclusion}
Their ingenious gating system allowed GRA to test, perhaps for the first time, quantum mechanical predictions for a single photon state. Interference is confirmed in the obvious way. The which-path predictions are also confirmed; the photon is detected in only one path. What we have shown though, is that a wave model (CIEM) can explain this result. It cannot therefore  be concluded that the detection of the photon on one path confirms particle behaviour. In a particle model, the photon takes one path at the beam-splitter and is detected in that path, whereas in our wave model the photon splits at the beam-splitter, is nonlocally absorbed, and is again detected in only one path. Since the  which-path measurement does not confirm particle behaviour, Bohr's principle of complementarity is also not confirmed, contrary to what is claimed by GRA. We conclude then, that GRA's experiments do not confirm complementarity. We may further add that if complementary is accepted, Wheeler's delayed-choice experiments lead to very strange conclusions: either history is changed at the time of measurement, or history is created at the time of measurement \cite{K05, BDH85}. CIEM, on the other hand, explains Wheeler's delayed-choice experiments in a unique and causal way.  

\section{Acknowledgements} 
I would like to thank Dr H.V. Mweene and Mr Y. Banda for proof reading my article. I would also like to thank the Dean of Natural Sciences, Dr S.F. Banda, for granting me a short study leave to complete this article.


\begin{thebibliography}{99}
\bibitem{T09} G.I. Taylor, Proc. Cambridge Philos. Soc. \underline{15}, 114 (1909); A.J. Dempster and H.F. Batho, Phys. Rev. \underline{30}, 644 (1927); L. Janossy and Z. Naray, Acta Phys. Hungaria \underline{7}, 403 (1967); H.M. Griffiths, Princeton University Senior Thesis (1963); G.T. Reynolds et al, Advances in electronics and electron physics  \underline{28B} (Academic Press, London, 1969); Y.P. Dontsov and A.I. Baz, Sov. Phys. JETP \underline{25}, 1 (1967); G.T. Reynolds, K. Spartalian and D.B. Scarl, Nuovo Cim. \underline{B61}, 355 (1969); P. Bozec, M. Cagnet and G. Roger, C.R. Acad. Sci. \underline{ 269}, 883 (1969); A. Grishaev et al, Sov. Phys. JETP \underline{32}, 16 (1969). 
\bibitem{GHOSE91} P. Ghose , D, Home and G.S. Agarwal, Phys. Lett. A \underline{153}, 403 (1991).
\bibitem{MIZ92} Y. Mizobuchi and Y. Ohtak\'{e}, Phys. Lett. A \underline{168}, 1 (1992).
\bibitem{UNNIK} C. S. Unnikrishnan and S. A. Murthy, Phys. Lett. A \underline{221}, 1 (1996).
\bibitem{GHOSE99} P. Ghose, {\it Testing Quantum Mechanics on a New Ground} (Cambridge Univ. Press, Cambridge, 1999).
\bibitem{BRIDA04}G. Brida,  M. Genovese, M. Gramegna, E. Predazzi, Phys. Lett. A \underline{328}, 313 (2004).
\bibitem{AFSHAR04} S. S. Afshar, http://irims.org/quant-ph/030503/.
\bibitem{WZ} W. K. Wootters and W.H. Zurek, Phys. Rev. D \underline{19}, 473 (1979)
\bibitem{GY}  D.M. Greenberger and A. Yasin, Phys. Lett A \underline{128}, 391(1988).
\bibitem{KPW} D. Home and P.N. Kaloyerou, J. Phys. A: Math. Gen. \underline{22}, 3253 (1989); P.N. Kaloyerou and H.R. Brown,  Physica B \underline{176}, 78 (1992); P.N. Kaloyerou, Found. Phys. \underline{22}, 1345 (1992).
\bibitem{KIM00} Y. Kim, R. Yu, S.P. Kulik, Y. Shih, M.O. Scully, Phys Rev. Lett. \underline{84}, 1 (2000)
\bibitem{WHR78}  J.A. Wheeler, in {\it Mathematical Foundations of Quantum Theory} edited by E.R. Marlow (Academic Press, New York, 1978), p. 9.
\bibitem{K05} P.N. Kaloyerou, Physica A \underline{355}, 297 (2005).
\bibitem{W26} G. Wentzel, Z. Physik \underline{40}, 574 (1926); L. Mandel, E.C.G. Sudarshan and E. Wolf, Proc. Phys. Soc. (London) \underline{84}, 435 (1964); W.E. Lamb Jr. and M.O. Scully in {\it Polarization: Mati\`{e}re Rayonnement}, (Presses Universitaires de France, Paris, 1969), p. 363. 
\bibitem{VW76} P.M. Mathews and K. Venkatesan, {\it A textbook of Quantum Mechanics} (Tata McGraw-Hill, New Delhi, 1976).
\bibitem{K85} P.N. Kaloyerou, Ph.D. Thesis (University of London, 1985).
\bibitem{K87} D. Bohm, B.J. Hiley, and P.N. Kaloyerou, Phys. Rep. \underline{144}, 349 (1987).
\bibitem{K94} P.N. Kaloyerou, Phys. Rep. \underline{244}, 287 (1994).
\bibitem{BR59A} N. Bohr in {\it Albert Einstein: Philosopher Scientist}, edited by P.A. Schilpp (Open Court, Lasalle,  Illinois, third edition, 1982) p. 201. 

\bibitem{JAM74}M. Jammer, {\it The Philosophy of Quantum Mechanics: The Interpretations of Quantum Mechanics in Historical Perspective} (John Wiley \& Sons, 1974).

\bibitem{BR28} N. Bohr at {\it Atti del Congresso Internazionale dei Fisici}, Como, 11-20 September 1927 (Zanichelli, Bologna, 1928), Vol. 2   p. 565; substance of the Como lecture is reprinted in Nature \underline{ 121}, 580 (1928) and in N. Bohr, {\it Atomic Theory and the Description of Nature} (Cambridge University Press, Cambridge, 1934) p. 52.

\bibitem{G86} P. Grangier, G. Roger and A. Aspect, EuroPhys. Lett. \underline{1}, 173 (1986).

\bibitem{AG86}A. Aspect in {Sixty-Two Years of Uncertainty}, edited by A.I. Miller (Plenum Press, New York, 1990) p. 45; A. Aspect and P. Grangier, Hyperfine Interactions \underline{37}, 3 (1987); A. Aspect, P. Granger, G. Roger, J. Optics (Paris). \underline{20}, 119 (1989). 

\bibitem{GHOSE93}P. Ghose, D. Home, M.N.S. Roy, Phys. Lett. A \underline{183}, 267 (1993).

\bibitem{GHOSE96}P. Ghose, Found. Phys. \underline{26}, 1441 (1996).

\bibitem{GHOSE01} P. Ghose, A.S. Majumdar, S. Guha and J. Sau, quant-ph/0102071 (2001).

\bibitem{KEMMER39}N. Kemmer, Proc. R. Soc. (London) \underline{A173}, 91 (1939).

\bibitem{E1905} A. Einstein, Ann. D. Physik \underline{17}, 132 (1905).

\bibitem{HVTHR} M. Genovese, Phys. Rep. \underline{413}, 319 (2005);  S. J. Belinfante, {\it A Survey of Hidden-Variables Theories} (Pergamon, Oxford, 1973); J. S. Bell, {\it Speakable and Unspeakable in Quantum Mechanics} (Cambridge University Press, Cambridge,  1987). 

\bibitem{SC2005} S. Colin, Ph.D. thesis (Universiteit Brussel, 2005), quant-ph/0301119.  

\bibitem{WS2005} W. Struyve, Ph.D thesis (Universitiet Gent, 2005), quant-ph/0506243.

\bibitem{AGR81} A. Aspect, P. Grangier, and G. Roger, Phys. Rev. Lett. \underline{47}, 460 (1981).

\bibitem{BJ89} B.H. Bransden and C.J. Joachain, {\it Quantum Mechanics} (Prentice Hall, London, 1989).

\bibitem{L73} R. Loudon, {\it The Quantum Theory of Light} (Oxford University Press, Oxford, 1973).

\bibitem{M76} L. Mandel in {\it Progress in Optics vol. 13}, ed. E. Wolf (North-Holland, Amsterdam, 1976).

\bibitem{BS73} I.N. Bronshtein and K.A. Semendyayev, {\it A Guide Book to Mathematics} (Springer-Verlag, New York, 1973).

\bibitem{C80} C.M. Caves, Phys. Rev. Lett. \underline{45}, 75 (1980).

\bibitem{OHM87} Z.Y. OU, C.K. Hong and L. Mandel, Opt. Commun. \underline{63}, 118 (1987). 

\bibitem{SZ97} M.O. Scully and M. S. Zubairy, {\it Quantum Optics} (Cambridge University Press, Cambridge, 1997).

\bibitem{CST} R.A. Campos, B.E.A. Saleh, and M.C. Teich, Phys. Rev. A \underline{40}, 1371 (1989).

\bibitem{F73} E.S. Fry, Phys. Rev. A \underline{8}, 1219 (1973).

\bibitem{BJD} J.D. Bjorken and S. D. Drell, {\it Relativistic Quantum Fields} (McGraw-Hill, New York, 1965).

\bibitem{LHR} L. H. Ryder, {\it Quantum Field Theory} (Cambridge University Press, Cambridge, second edition, 1996 ).

\bibitem{SHFF68} L.I. Schiff, {\it Quantum Mechanics} (McGraw-Hill Kogakusha, third edition, 1968).

\bibitem{QE} U. Mohrhoff, Am. J. Phys.  \underline{64}, 1468 (1996); B. -G. Englert, M.O. Scully and H. Walther, Am. J. Phys.  \underline{67}, 325 (1999); U. Mohrhoff, Am. J. Phys.  \underline{67}, 330 (1999); A.G. Zajonic, L.J. Wang, X.Y. Zou, L. Mandel, Nature (London) \underline{353}, 507 (1991); P.G. Kwiat, A. M. Steinberg  and R.Y. Chiao, Phys. Rev. A \underline{45}, 7729 (1992); T.J. Herzog, P.G. Kwiat, H. Weinfurter, and A. Zeilinger, Phys. Rev. Lett. \underline{75}, 3034 (1995); T.B. Pittman, D.V. Strekalov, A. Migdall, M.H. Rubin, A.V. Sergienko, and Y.H. Shih, Phys. Rev. Lett. \underline{77}, 1917 (1996)

\bibitem{BDH85}  D. Bohm,  C. Dewdney, and B.J. Hiley,  Nature \underline{315}, 294 (1985).





\end{thebibliography}
\end{document}